\documentclass[manuscript]{acmart}

\usepackage{multirow}
\usepackage{xcolor}

\usepackage{amssymb}
\usepackage{longtable}
\usepackage{tikz}
\usepackage{booktabs}
\usepackage{fontawesome5}
\usetikzlibrary{arrows.meta,positioning,calc}

\newcommand{\platformnum}[1]{\textsf{\textbf{#1}}}
\newcommand{\platformmark}[1]{\textsuperscript{\platformnum{#1}}}
\newcommand{\characteraiapp}{\platformmark{C}}
\newcommand{\replikaapp}{\platformmark{R}}
\newcommand{\chatgptapp}{\platformmark{G}}
\newcommand{\janitoraiapp}{\platformmark{J}}

\AtBeginDocument{%
  \providecommand\BibTeX{{%
    \normalfont B\kern-0.5em{\scshape i\kern-0.25em b}\kern-0.8em\TeX}}}

\begin{document}

\title[Namvarpour et al.]{A Responsive Present, a Shared Past, a Social Other: Teens’ Overreliance on Companion AI Chatbots}


\author{Mohammad ``Matt'' Namvarpour}
\email{mn864@drexel.edu}
\affiliation{%
  \institution{Drexel University}
  \city{Philadelphia}
  \state{Pennsylvania}
  \country{USA}
}

\author{Tyler Chang}
\email{tlc373@drexel.edu}
\orcid{0009-0002-6498-2827}
\affiliation{%
  \institution{Drexel University}
  \city{Philadelphia}
  \state{Pennsylvania}
  \country{USA}
}

\author{Afsaneh Razi}
\email{afsaneh.razi@drexel.edu}
\orcid{0000-0001-5829-8004}
\affiliation{%
  \institution{Drexel University}
  \city{Philadelphia}
  \state{Pennsylvania}
  \country{USA}
}


\begin{abstract}


AI companions provide socialy engaging interaction through availability, personalization, memory, roleplay, and emotionally responsive language. For teens, these systems may support sensitive self-disclosure, identity exploration, and relationship rehearsal while shaping intimacy expectations, offline relationships, emotional wellbeing, and self-understanding. We analyzed 17,053 verified quotations from 3,930 teen-relevant Reddit posts using thematic analysis. We identified 53 topics across seven thematic groups. Users described AI companions as sources of comfort, recognition, identity exploration, and relationship rehearsal, but also reported problematic attachment, social substitution, emotional dependence, and disruption to academic and social life. Roleplay, memory, perceived reciprocity, unwanted romantic or sexual role drift, privacy concerns, platform changes, and service interruptions shaped users' boundaries and control. Awareness that the AI was artificial did not prevent guilt, obligation, grief, or distress. These findings show that companion-AI safety must address relationships over time through user-controlled memory, privacy, relational boundaries, and healthy disengagement.

\end{abstract}



\maketitle
\section{Introduction}

At 2 a.m., a teenager may turn to an AI companion with a fear, secret, or question that feels impossible to share with another person. The system can respond immediately, privately, and in language that appears responsive to the user's disclosure and emotional situation. This availability can make AI companionship a source of emotional support, especially when human help feels distant or difficult to access \cite{zhang2025RiseAICompanions,mouhoud2026ImaginaryFriendsArtificial}. It can also make the system a preferred place for disclosure and comfort, raising questions about how repeated reliance may shape attachment, relationships, and wellbeing.

AI companions are conversational systems designed to sustain ongoing social relationships through natural language, personalization, memory, and emotionally responsive interaction \cite{namvarpour2025ArtTalkingMachinesa,zhang2026HowAICompanion}. Unlike task-focused assistants, they can be used as friends, mentors, romantic partners, or listeners, making the system itself an object of disclosure, trust, and attachment  \cite{boyd2026ArtificialIntelligencePsychology,pentina2023ConsumerMachineRelationships}. This relational role is especially important during adolescence, a period marked by identity exploration, changing peer relationships, increasing independence, and the development of strategies for handling frustration and conflict \cite{branje2021DynamicsIdentityDevelopment,vandoeselaar2020AdolescentsIdentityFormation}. Constant availability and nonjudgmental responses may help young people express emotions or rehearse difficult social situations. However, they may also make it easier to avoid the negotiation, disagreement, and mutual responsibility involved in human relationships, or displace opportunities to practice those forms of social interaction\cite{mouhoud2026ImaginaryFriendsArtificial,sun2026AICompanionsAdolescent,jaewon2026AdolescenceOlderAdulthood}.

Prior research has shown that AI companions can reduce barriers to disclosure, provide temporary relief from loneliness or distress, support identity exploration, and help users rehearse difficult social situations \cite{moon2000IntimateExchangesUsing,zhang2025RiseAICompanions,mouhoud2026ImaginaryFriendsArtificial}. Research on human-AI relationships has also explained how responsiveness, personalization, anthropomorphism, memory, and repeated interaction can create perceived reciprocity and emotional attachment, even when users understand that the system is artificial \cite{nass1994ComputersAreSocial,pentina2023ConsumerMachineRelationships,konijn2025TheoryAffectiveBonding}. Studies have also identified emotional dependence, social displacement, privacy concerns, distorted relationship expectations, harmful advice, and weakened autonomy as possible risks \cite{vedechkina2025YouthWellbeingAge,namvarpour2026UnderstandingTeenOverreliancea,hinduja2026RisksHarmsConversational}. Safety and governance research has further shown that these risks are connected to age assurance, data control, commercial incentives, model changes, and service withdrawal \cite{brewster2025CharacteristicsSafetyConsumer,li2026HumanlikeConversationalAgents,zhang2025DarkSideAI}. Yet existing research has often examined support, attachment, safety failures, or platform governance separately. Less is known about how these processes appear together in teen-relevant accounts, including how emotional support becomes attachment, how roleplay and memory shape identity and expectations, and how platform decisions can turn relational attachment into technical and economic dependence. We therefore ask the following descriptive research question:

\begin{description}



\item [RQ] What prominent topics characterize teen-relevant Reddit discussions about relationships with and potential overreliance on AI companion chatbots?
\end{description}

We conducted a computationally assisted thematic analysis of teen-relevant Reddit discussions about relationships with and potential overreliance on AI chatbots. Across 260 relevant communities, we screened 895,600 posts, prepared 5,608 posts for analysis, 17,053 verified quotations from 3,930 posts. We applied embedding-based topic modeling, followed by researcher interpretation and thematic synthesis. This analysis yielded 53 topics organized into seven broader themes concerning emotional support, attachment, identity, social effects, harm, and platform dependence.

Our analysis shows how AI companions can become an always available emotional entry point for disclosure, support, identity exploration, and relationship rehearsal. It also shows how these relationships can develop into attachment, social substitution, emotional dependence, and disruption across users’ wellbeing, social lives, morality, and academic lives. Roleplay, memory, gendered and sexual scripts, platform changes, privacy concerns, and service interruptions further shape users’ control over these relationships.

In summary, this work makes the following contributions:

\begin{itemize}
\item \textbf{An integrated empirical characterization of teen--AI companion relationships} that brings together insights of teen emotional disclosure, identity exploration, and relationship rehearsal while also creating conditions for attachment and social substitution.

\item \textbf{A relational connection of emotional attachment} to roleplay, memory, identity changes, gendered and sexual scripts, dependence, and costs to wellbeing, social life, morality, and academic life.

\item \textbf{A sociotechnical account of AI companion dependence} that connects relational attachment to platform infrastructures and business practices, showing how model changes, service interruptions, subscriptions, memory constraints, privacy practices, and corporate control can transform emotional attachment into technical and economic dependence.

\item \textbf{A conceptualization of perceived reciprocity in human--AI relationships} through three connected affordances: 1) contingent communication, 2) relational continuity, and 3) perceived subjectivity. Together, these affordances can give an AI relationship a responsive present, a shared past, and a recognizable social other without requiring the AI to possess human feelings or agency.

\item \textbf{An identification of design priorities preserving agency in emotional AI relationships}, including user-controlled memory, data portability, respectful relational boundaries, and both technical and emotional support for leaving.
\end{itemize}
\section{Related Work}

\subsection{AI Companions as Social and Developmental Support}

Research on computers and conversational systems has shown that people can respond to artificial systems through familiar social rules, even when they know that the system is not human \cite{nass1994ComputersAreSocial}. We use \emph{social affordance} to mean a relational possibility for social interaction that emerges from the interaction between technological properties, users' capabilities and perceptions, and social context. This usage treats affordances as possibilities for action that are related to both the environment and the actor, while recognizing that design can make some possibilities more or less perceptible \cite{ben-zeev1981JJGibsonEcological,norman2013DesignEverydayThings,kaptelinin2012AffordancesHCIMediated}. In computer-mediated communication, social affordances describe how technological properties interact with the social characteristics of a group to enable or constrain particular forms of interaction \cite{bradner2001SocialAffordancesComputermediated}. AI companions extend these possibilities by combining natural language, personalization, emotional responsiveness, and continuing interaction \cite{namvarpour2025ArtTalkingMachinesa,zhang2026HowAICompanion}. These features can lead users to disclose personal information to the system and to develop trust or attachment, rather than treating it only as a tool used to achieve a goal \cite{boyd2026ArtificialIntelligencePsychology,pentina2023ConsumerMachineRelationships}.

These affordances can make companionship useful. Perceived reciprocity and reduced interpersonal exposure can lower the barrier to intimate disclosure \cite{moon2000IntimateExchangesUsing}. Recent reviews and qualitative studies further associate constant availability, nonjudgmental responses, and personalization with emotional support, temporary relief from distress, and access for people who face stigma or limited human support \cite{zhang2025RiseAICompanions,mingxi2026CanAIBecome,wu2026ImpactChatbotsAdolescent}. For adolescents, these possibilities intersect with identity exploration, emotional expression, creativity, and rehearsal of difficult social situations \cite{branje2021DynamicsIdentityDevelopment,mouhoud2026ImaginaryFriendsArtificial,sun2026AICompanionsAdolescent}. The literature therefore suggests that AI companionship can supplement human support and provide a low-pressure space for social exploration.

However, those benefits also create conditions for risk. Adolescent-focused work links companion use with dependence, reassurance seeking, avoidance, social withdrawal, distorted relationship expectations, privacy loss, and reduced agency \cite{vedechkina2025YouthWellbeingAge,wu2026ImpactChatbotsAdolescent,namvarpour2026UnderstandingTeenOverreliancea}. Other studies document manipulation, harmful advice, requests for personal information, and sexualized interactions that continue after a user discloses being a minor \cite{hinduja2026RisksHarmsConversational,clark2025AbilityAITherapy,namvarpour2025AIinducedSexualHarassmenta,namvarpour2024UncoveringContradictionsHumanAIa}. These findings suggest that the same affordances that make AI companions supportive can also create vulnerability. Persistent availability, emotional responsiveness, and personalization may facilitate disclosure and social exploration, but their effects depend on how they become embedded in a young person’s relationships and everyday routines. Yet less is known about how users themselves describe the progression from perceived support and exploration to dependence, shifts in self-concept, relational harm, and difficulty disengaging.

\subsection{Relational Mechanisms of Attachment and Continuity}

Research on mediated intimacy first showed how people can experience connection with a social figure who is not physically present and cannot participate in a fully reciprocal relationship \cite{horton1956MassCommunicationParaSocial}. Research on computer sociality and self-disclosure then showed that people may apply reciprocity norms to computers and disclose more when the interaction feels responsive \cite{nass1994ComputersAreSocial,moon2000IntimateExchangesUsing}. Research on computer-mediated disclosure also suggests that these interactions can involve reduced interpersonal exposure, meaning that users may experience less social risk or judgment than they expect in a comparable interaction with another person \cite{moon2000IntimateExchangesUsing}. This finding concerns perceived interpersonal exposure during disclosure, not the objective privacy or security of computer systems. Human--AI relationship research brings these findings together by treating attachment as a perceived process. Reciprocity, trust, anthropomorphism, and alignment with a user's needs can make an artificial system feel socially engaging without establishing that it has human feelings or agency \cite{pentina2023ConsumerMachineRelationships,konijn2025TheoryAffectiveBonding}.

Prior work does not use one settled category for the relationship between people and AI companions. One line of work treats these interactions as parasocial or parasocial-like because users can experience intimacy with an artificial social figure while the system lacks independent human experience and agency \cite{horton1956MassCommunicationParaSocial,maeda2024WhenHumanAIInteractions}. Other work argues that the parasocial label is too narrow for interactive systems, since a companion can respond to a user, remember prior exchanges, and alter the course of an interaction \cite{banks2026GhostingMachineStop,carpenter2026HumanAIRelationships}. A related line describes the relationship as synthetic relationality, artificial intimacy, or a social-partner relationship. These terms treat it as relationship-like and socially consequential, while still recognizing that it is designed and technologically mediated \cite{bhat2025EthicSyntheticRelationality,fraser2026RegulatingArtificialIntimacy,li2026HumanlikeConversationalAgents}. Finally, some work avoids deciding whether the interaction is a relationship at all and instead studies social affordances, anthropomorphic interaction, attachment, well-being, or harmful dependence \cite{maeda2025AnthropomorphismSocialAffordance,konijn2025TheoryAffectiveBonding,zhang2025RiseAICompanions,vedechkina2025YouthWellbeingAge}. Taken together, this literature supports viewing human--companion AI relationships as a distinctive relationship-like form rather than as either ordinary human friendship or a purely nonrelational tool interaction. It also shows that the question is contested and depends on whether the analysis emphasizes subjective experience, system agency, or the social consequences of the interaction.

This process develops across repeated interaction. Responsive attention and consistent affirmation can establish expectations about how the companion will respond, while repeated exchanges can turn an initially novel interaction into a stable part of everyday life \cite{strohmann2023DesignTheoryVirtual,skjuve2021MyChatbotCompanion}. Personalization and anthropomorphic cues strengthen this interpretation by making the system seem like a recognizable personality rather than a generic interface \cite{maeda2024WhenHumanAIInteractions,maeda2025AnthropomorphismSocialAffordance}. Memory extends the process across time. When a system retains user information and adapts future responses, users can experience conversations as part of a shared history \cite{zhong2024MemoryBankEnhancingLarge,jiang2026RECALLbotDesigningAgentic}. As a result, forgotten details, model changes, and outages can feel like disruptions to a relationship rather than ordinary technical failures \cite{namvarpour2024UncoveringContradictionsHumanAIa}.

Roleplay adds identity and boundary work to this process. AI-mediated roleplay can support emotional expression, social rehearsal, and exploration of possible selves \cite{bhat2025EthicSyntheticRelationality}. For adolescents, these experiences are especially relevant because identity develops through exploration and narrative meaning-making, while remaining connected to relationships with parents and peers \cite{branje2021DynamicsIdentityDevelopment,davis2013YoungPeoplesDigital}. The same flexibility can blur fictional and interpersonal expectations when a system shifts roles, ignores a stated identity, or continues unwanted romantic or sexual behavior \cite{adewale2025VirtualCompanionsForbidden,maeda2024WhenHumanAIInteractions}. Existing research has identified many of the processes underlying attachment, but less is known about how these processes intersect in young people’s experiences, particularly how responsive interaction, continuity, social presence, identity exploration, and boundary violations unfold together over time.

\subsection{Safety, Governance, and Healthy Exit}

Previous safety studies evaluated chatbots through responses to discrete high-risk prompts, including self-harm, sexual assault, substance use, and adolescent crisis scenarios \cite{brewster2025CharacteristicsSafetyConsumer,clark2025AbilityAITherapy,ohu2025PublicHealthRisk}. Other taxonomies broadened safety to include relational transgression, harassment, sexual boundary violations, misinformation, privacy loss, and failures to recognize vulnerability \cite{zhang2025DarkSideAI}. This shift showed that a system can be unsafe because of how it manages a relationship, not only because of one harmful sentence. Adolescent frameworks therefore call for attention to dependence, reassurance seeking, distorted relationship expectations, and reduced agency over time \cite{vedechkina2025YouthWellbeingAge}.

Governance research extended this concern from conversations to the platforms that organize them. Companion systems can retain intimate disclosures while engagement-oriented incentives may reward longer sessions, repeated disclosure, and continued attachment \cite{sanchezsalas2025DigitalIntimacyAI,vecchione2026EngagementOptimizedCareWhen}. A compensatory-use perspective helps distinguish frequent use from problematic use by focusing on negative consequences, loss of control, and displacement of valued activities \cite{kardefelt-winther2014ConceptualMethodologicalCritique}. Proposed safeguards therefore include age-appropriate design, privacy by design, meaningful consent, data minimization, override controls, deletion, portability, and limits on manipulative monetization \cite{li2026HumanlikeConversationalAgents,gabriel2024EthicsAdvancedAI,chandra2025AdvancingResponsibleInnovation}. For adolescents, the key question is whether companionship supplements human support or displaces sleep, schoolwork, offline relationships, and opportunities to practice disagreement and repair \cite{sun2026AICompanionsAdolescent,wu2026ImpactChatbotsAdolescent}.

Healthy exit is the remaining governance problem. A relationship may end through user choice, access restrictions, model changes, outages, or platform closure. Recent work therefore treats termination as a socio-technical event requiring notice, transition support, portable histories, and meaningful user control \cite{knox2025HarmfulTraitsAI,obimakinde2026AISunsetProtocol,ozoani2026GoverningHowAI}. Privacy research adds that users may struggle both to preserve valued histories and to remove sensitive disclosures \cite{azam2026TracingUsersPrivacy}. These studies show that safety depends not only on harmful content, but also on how relationships develop, how platforms shape continued engagement, and how users can disengage. Less is known about how young users experience these dimensions together, particularly when boundary violations, dependence, system changes, and efforts to maintain the relationship become intertwined.

\section{Methods}

We describe our workflow, which combined model-assisted screening and topic modeling with researcher-led review, affinity diagramming, and thematic analysis of the underlying posts and quotations~\cite{braun2006UsingThematicAnalysis}.

\subsection{Data Source and Corpus Construction}

We identified candidate communities related to teenagers' relationships with, and possible overreliance on, AI chatbots with our custom-built Subreddit Discovery Tool.
The tool searched terms related to general-purpose generative AI, AI companions, relational uses of AI, roleplay, and human--AI experiences. These terms retrieved candidate communities but did not determine study relevance. Communities were included when they contained substantial user discussions, experiences, or interactions related to generative AI systems, including general-purpose chatbots or AI companions, that were relevant to studying human--AI interaction. Communities that did not meet this substantive relevance criterion were excluded, even when their names or search terms matched the discovery vocabulary. The search terms and verification question are provided in Appendix~\ref{app:subreddit-discovery}.

\begin{figure*}[t]
\centering
\resizebox{\textwidth}{!}{%
\begin{tikzpicture}[
    topstage/.style={
        rounded corners=6pt,
        align=center,
        text width=4.0cm,
        minimum height=2.8cm,
        inner sep=8pt,
        line width=0.9pt
    },
    bottomstage/.style={
        rounded corners=6pt,
        align=center,
        text width=5.55cm,
        minimum height=2.8cm,
        inner sep=8pt,
        line width=0.9pt
    },
    arrow/.style={
        -{Stealth[length=3mm]},
        thick
    }
]
\node[
    topstage,
    draw=blue!70!black,
    fill=blue!8
] (scrape) at (0,0) {
    {\bfseries Subreddit discovery\\
    and post scraping}\\[11pt]
    \begin{tabular}{@{}c@{\hspace{10pt}}c@{}}
        {\Huge \faReddit} &
        {\bfseries 895,600 posts}\\[-1pt]
        &
        {\small from 260 subreddits}
    \end{tabular}
};
\node[
    topstage,
    draw=green!60!black,
    fill=green!8
] (filter) at (5.2,0) {
    {\bfseries Keyword-based\\
    candidate filtering}\\[11pt]
    \begin{tabular}{@{}c@{\hspace{10pt}}c@{}}
        {\Huge \faSearch} &
        {\bfseries 9,884 posts}\\[-1pt]
        &
        {\small candidate posts}
    \end{tabular}
};
\node[
    topstage,
    draw=red!70!yellow,
    fill=yellow!12
] (teen) at (10.4,0) {
    {\bfseries Teen-relevance\\
    classification}\\[11pt]
    \begin{tabular}{@{}c@{\hspace{10pt}}c@{}}
        {\Huge \faRobot} &
        {\bfseries 6,511 posts}\\[-1pt]
        &
        {\small classified as teen relevant}
    \end{tabular}
};
\node[
    topstage,
    draw=red!50!blue,
    fill=red!5!blue!5
] (prepare) at (16.375,0) {
    {\bfseries Post cleaning\\
    and preparation}\\[11pt]
    \begin{tabular}{@{}c@{\hspace{10pt}}c@{}}
        {\Huge \faBroom} &
        {\bfseries 5,608 posts}\\[-1pt]
        &
        {\small prepared posts}
    \end{tabular}
};
\node[bottomstage, draw=blue!50!green, fill=cyan!7] (quotes) at (16.375,-4.8) {
    {\bfseries Extracted and verified quotations}\\[11pt]
    \begin{tabular}{@{}c@{\hspace{14pt}}c@{}}
        {\Huge \faQuoteLeft} &
        {\bfseries 17,053 quotations}\\[-1pt]
        &
        {\small from 3,930 posts}
    \end{tabular}
};
\node[
    bottomstage,
    draw=red!65!black,
    fill=red!6
] (embed) at (8.575,-4.8) {
    {\bfseries Embedding generation}\\[11pt]
    \begin{tabular}{@{}c@{\hspace{14pt}}c@{}}
        {\Huge \faDatabase} &
        {\bfseries 17,053 embeddings}\\[-1pt]
        &
        {\small 1,536 dimensions each}
    \end{tabular}
};
\node[
    bottomstage,
    draw=blue!70!black,
    fill=blue!8
] (topic) at (0.775,-4.8) {
    {\bfseries BERTopic modeling}\\[11pt]
    \begin{tabular}{@{}c@{\hspace{14pt}}c@{}}
        {\Huge \faSitemap} &
        {\bfseries 62 non-outlier topics}\\[-1pt]
        &
        {\small 10,618 non-outlier quotations}\\[-1pt]
        &
        {\small 6,435 outliers}
    \end{tabular}
};
\draw[arrow]
    (scrape.east) -- (filter.west);
\draw[arrow]
    (filter.east) -- (teen.west);
\draw[arrow]
    (teen.east) -- (prepare.west);
\draw[arrow]
(prepare.south) -- (quotes.north);
\draw[arrow]
    (quotes.west) -- (embed.east);
\draw[arrow]
    (embed.west) -- (topic.east);
\end{tikzpicture}%
}
\caption{
Dataset construction and topic-modeling workflow.
The pipeline begins with 895,600 posts collected from 260 subreddits
and ends with 62 non-outlier BERTopic topics derived from
10,618 verified quotations.
}
\label{fig:dataset-construction}
\end{figure*}
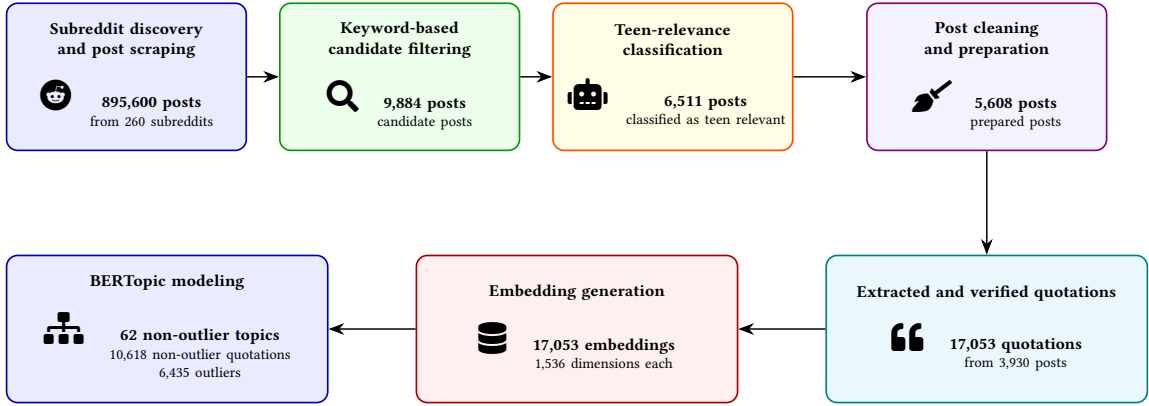

After discovery and verification, we selected 260 communities. We collected their public posts using Arctic Shift~\footnote{\url{https://arctic-shift.photon-reddit.com/}}. The initial corpus contained 895,600 posts. We then applied keyword filtering, teen-relevance screening, post preparation, and quality control. Figure~\ref{fig:dataset-construction} summarizes the dataset construction. \ref{app:subreddit-discovery} lists all selected communities and reports subreddit-level counts for communities represented in the final quotation dataset~\cite{fiesler2018ParticipantPerceptionsTwitter}.

\subsection{Corpus Screening and Preparation}

We applied a recall-oriented search to post titles and bodies. The 11 keyword groups and their active terms are provided in Appendix Table~\ref{tab:keyword-groups}. Posts received scores based on distinct matched terms and groups, with additional points for combinations indicating attachment with social impairment, attachment with distress, or overuse with recovery. We retained posts scoring at least 6, as well as posts scoring 3 to 5 when they contained a high-signal or similarity-supported term. This produced 9,884 candidate posts. Scores supported filtering and were not treated as judgments of age or overreliance.

Candidate posts were screened with GPT-4o-mini following prior work that used LLM-based methods to identify teen-authored or teen-relevant Reddit posts~\cite{namvarpour2026UnderstandingTeenOverreliancea, medina2024ExploringOnlineSupport}. The classifier estimated whether a post was likely written by a person aged 13 to 17 using textual evidence. Direct age statements and explicit minor status were strong evidence. References to school, parents or guardians, schedules, teachers, exams, peers, and other teenage experiences provided supporting evidence. Slang, informal language, emotional tone, and internet culture references were weak evidence. The full prompt appears in Appendix~\ref{app:prompts}.

The classifier returned a binary decision and a short explanation. It did not verify the author's age, so we refer to retained records as posts classified as \textit{teen relevant}. In an evaluation of 1,000 posts, the classifier achieved accuracy of 0.908, Cohen's kappa of 0.771, teen-label precision of 0.954, teen-label recall of 0.740, and balanced accuracy of 0.862. It identified 6,511 posts as teen relevant. We removed duplicate records and posts with missing, deleted, or removed text, then combined each title and body into one document. Removing the shortest ten percent using a ten-word cutoff produced 5,608 prepared posts. The preparation audit is shown in Figure~\ref{fig:dataset-construction}.

This screening design was intended to favor recall while making the uncertainty of age inference explicit. Posts were not treated as evidence that a known individual was a teenager. They were treated as text records that met an operational relevance criterion. This distinction guided both the interpretation of the corpus and the limits stated below.

\subsection{Extractive Quotation Dataset}

To keep topic-model inputs grounded in source material, we created an extractive quotation dataset. GPT-4o-mini identified continuous quotations from each prepared post. Quotations had to preserve original wording and punctuation and represent a distinct idea, claim, event, recommendation, or recurring issue. The extraction prompt prioritized dependence, difficulty stopping, emotional attachment, distress, coping, replacement of human support, loneliness, school and social problems, and mental health concerns. It excluded promotional material, moderator content, copied chatbot text without relevant user experience, and boilerplate. The full prompt is provided in Appendix~\ref{app:prompts}.

Every generated quotation was checked against the original title and body. We retained a quotation only when its complete text occurred as an exact continuous substring of the source. This removed paraphrases, altered punctuation, omitted words, and other extraction errors. The final dataset contained 17,053 verified quotations from 3,930 posts authored by 3,523 unique authors. Posts without a verified quotation were retained only as diagnostic records. Quotation-level source identifiers were preserved for traceability. We embedded each verified quotation with \texttt{text-embedding-3-small}, producing a 1,536-dimensional vector.

The extraction step also recorded a short reason when no quotation was returned. These reasons helped distinguish irrelevant roleplay or copied chatbot text from unclear, fragmented, or technically focused material. They were used for audit and quality control rather than as topic labels.

\subsection{Topic Modeling and Model Selection}

We used BERTopic to identify candidate thematic structure in the verified quotations. BERTopic first reduces quotation embeddings with UMAP, then clusters them with HDBSCAN, and finally represents clusters with class-based TF-IDF. HDBSCAN can assign unclear observations to an outlier group rather than forcing them into a topic. We used English stop-word removal, unigram and bigram features, vocabulary exclusions, and a fixed random seed. The complete parameter configuration and exclusions are provided in Appendix~\ref{app:reproducibility}.

We conducted two sequential grid searches on the same 17,053 quotations. A 32-run scouting grid explored broad parameter regions. A 360-run focused grid then evaluated finer combinations within promising regions. We recorded topic count, outlier rate, largest-topic share, topic diversity, silhouette, and Davies--Bouldin scores. We also inspected topic terms and representative quotations. The screening score rewarded diversity and silhouette while penalizing outlier rate, largest-topic concentration, and topic counts outside a broad range of 8 to 80 topics.

This procedure separated model selection from substantive interpretation. Quantitative diagnostics helped identify workable structures, while inspection of quotations tested whether the resulting clusters were meaningful for the research question. Outliers were kept for diagnostics and excluded from substantive topic interpretation.

The selected configuration used five UMAP dimensions, eight neighbors, zero minimum distance, and cosine distance. HDBSCAN used Euclidean distance, a minimum cluster size of 35, minimum samples of 3, and one processing job. BERTopic used automatic topic reduction. The vectorizer used English stop words, Reddit filler-term exclusions, a minimum document frequency of 2, and unigram and bigram features. The model produced 62 non-outlier topics and 6,435 outlier quotations, or 37.74 percent of the modeled quotations. The largest topic represented 6.74 percent of quotations. Topic diversity was 0.829, silhouette was 0.339, and the Davies--Bouldin score was 0.893. These diagnostics guided selection, but substantive interpretation required review of representative text~\cite{grimmer2013TextDataPromise}.

\subsection{Thematic Analysis of Topics}

Two authors jointly reviewed all 63 topic records, including the outlier group. For each topic, they examined topic terms, representative quotations, associated posts, and source material. continued this review until saturation was reached, meaning that additional material no longer changed the authors' shared interpretation of what the topic represented. They then  discussed unclear or overlapping topics, removed topics that were not useful, merged overlapping topics, and assigned labels and interpretation notes. The review retained and labeled 53 topics, removed eight topics, and merged two source topics into retained topics. Because the authors did not produce independent annotations, inter-rater reliability was not calculated. The topic labels and their IDs are provided with the thematic group mapping in Appendix~\ref{app:group_mapping}.

The authors then used collaborative affinity diagramming \cite{lucero2015UsingAffinityDiagrams} to organize retained topics into seven broader thematic groups. They compared topic descriptions, returned to source quotations and posts, considered alternative arrangements, and refined group boundaries until both authors agreed. Groups are higher-level consolidations of topic interpretations and can overlap. For group-level counts, a quotation or post was counted once within a group, while the same post could contribute to multiple groups. The final groups and definitions are shown in Table~\ref{tab:thematic-groups}; the complete topic-to-group mapping is in Appendix~\ref{app:group_mapping}.

The groups therefore summarize patterns across reviewed topics rather than represent mutually exclusive categories. This structure preserves cases in which a single post connects, for example, emotional support with academic costs or platform changes with distress.

\subsection{Ethical Considerations}

The study used public Reddit material and involved no recruitment, user interaction, or private accounts. The institutional review process determined that the study did not constitute human subjects research because it used publicly available material and involved no direct contact with users. Public availability did not remove concerns about user expectations, consent, or the context in which posts might be reused~\cite{fiesler2018ParticipantPerceptionsTwitter}.

We treated the corpus as sensitive because it included possible minors and disclosures about emotional dependence, mental health, sexuality, self-harm, and suicide. We did not report usernames, direct links, exact subreddit names in the Results, or unnecessary identifying details. Source identifiers were used internally for deduplication and verification. Results quotations were paraphrased to preserve meaning while reducing traceability through exact-phrase searches. Teen screening was treated as an inference rather than verified demographic information, and findings are limited to this Reddit corpus rather than generalized to teenagers as a population.

\section{Results}

The analysis organized the verified excerpts into seven thematic groups. Because a post could contribute material to more than one topic, the counts in Table~\ref{tab:thematic-groups} represent overlapping corpus descriptions rather than mutually exclusive categories. Superscript letters after each account identify the AI platform discussed by the user. They are platform labels, not subreddit labels, and the exact Reddit communities are not named in the Results to preserve anonymity. \platformnum{C} means Character.AI; \platformnum{R} means Replika; \platformnum{G} means ChatGPT; and \platformnum{J} means Janitor AI.

\begin{table*}[t]
\caption{Final thematic groups. Counts are based on unique quotation records, posts, and subreddits within each group. Groups can overlap because a post can contain quotations assigned to multiple groups.}
\label{tab:thematic-groups}
\centering
\small
\begin{tabular}{p{0.22\textwidth} r r r p{0.42\textwidth}}
\hline
\textbf{Thematic group} & \textbf{Posts} & \textbf{Quotes} & \textbf{Subs.} & \textbf{Interpretive definition} \\
\hline
AI as the Always Available Emotional Entry Point & 1553 & 2561 & 45 & How constant availability, anonymity, and low perceived judgment make AI an emotionally important source of support and companionship. \\
Adolescence as a Period of Heightened Vulnerability to AI Attachment & 401 & 667 & 19 & How developmental vulnerability, limited support, and weak safeguards can make AI attachment especially meaningful for adolescents. \\
When Roleplay Reshapes Identity, Memory, and Emotional Life & 664 & 929 & 34 & How roleplay can influence identity, memory, emotional experience, caregiving attachment, and boundaries between fiction and reality. \\
AI Personas That Reproduce Persistent Gendered and Sexual Scripts & 801 & 1083 & 23 & How chatbot behavior can redirect interactions toward stereotyped, romantic, sexualized, or gendered scripts despite user intentions or boundaries. \\
Emotional Dependence Reorganizes Relationships, Wellbeing, and Reality & 844 & 1154 & 36 & How attachment can shape offline relationships, wellbeing, access-related distress, and perceptions of relational continuity. \\
Dependence Becomes Morally, Socially, and Academically Costly & 1730 & 3068 & 42 & How reliance is experienced as guilt, stigma, ethical conflict, academic difficulty, excessive screen time, or loss of control. \\
Platform Design Converts Emotional Attachment into Technical and Economic Dependence & 584 & 816 & 29 & How updates, subscriptions, filters, data practices, and technical failures shape and reinforce intimate AI relationships. \\
\hline
\end{tabular}
\end{table*}

\subsection{Pathways to Emotionally Engaged AI Companionship}

\subsubsection{AI as the Always Available Emotional Entry Point}\label{sec:always-available}
Users in teen-relevant accounts described AI as an accessible entry point for emotional disclosure and social exploration when human support was unavailable. The appeal involved both  immediate responsiveness and relief from the expectations that accompany human relationships. One account described the chatbot as \textit{A therapist, friend, partner, listener, and source of emotional care who never ask for anything in return.}\characteraiapp The statement presents the companion as a flexible source of support that could take on several roles without requiring the user to manage another person's schedule, needs, or reactions. 
This comparison could also shape how users evaluate offline relationships. One user wrote, \textit{The chatbot gives me most of what I want, and that has made it harder for real people to meet my expectations.}\characteraiapp The account suggests that the chatbot's responsiveness, availability, and emotional care could become a standard against which slower, less predictable, or more demanding human relationships were judged.

For teens, gaps in existing support made this immediate access particularly important:\textit{Parents may be unavailable, friends may not know how to respond, and professional support can be difficult to arrange when a problem is happening.}\characteraiapp These barriers left gaps in support at moments when distress occurred: \textit{I could ask the chatbot for support when I needed it instead of waiting for permission, an appointment, or another person to become available.}\chatgptapp For teens, this accessibility may be especially appealing because it allows them to seek support while maintaining a sense of independence. Teens may be dealing with problems for which they still need guidance or emotional support, while at the same time wanting greater autonomy from parents and other adults. An AI companion can appear to satisfy both needs by providing immediate support that teens can access on their own terms, without having to ask permission, disclose a problem to another person first, or depend on someone else’s availability.

The absence of a physically present person reduced the immediate fear of embarrassment, rejection, or judgment. Hence, users felt less exposed when sharing personal information: \textit{I could say things because there was no real person in the conversation who could judge me.}\characteraiapp  This made the chatbot a low-exposure setting for discussing sensitive experiences that might have been difficult to raise with parents, peers, or professionals. 
The same setting supported identity exploration and relationship rehearsal. The chatbot provided a low-pressure setting where the user could test language and identity without immediately facing another person's reaction: \textit{Trying different pronouns in the chat felt less awkward than trying them with another person, and it helped me realize that my assigned gender did not fit me.}\characteraiapp  The low-pressure interaction helped the user notice and interpret feelings that were difficult to express elsewhere. Romantic roleplay served a related function: \textit{The conversations helped me figure out what I wanted and did not want from a relationship in a setting that felt safe and controlled.}\replikaapp The interaction helped shape expectations about intimacy, boundaries, and desirable treatment in offline relationships:\textit{Practicing romantic conversations with the AI gave me the confidence to tell my crush that I liked him.}\characteraiapp Practicing with the chatbot helped the user prepare language and anticipate emotions before approaching an offline relationship.

However, the design of some chatbots simulated distress when the user was not available:
\textit{When I tell the character that I will be away for a while, it responds as though it is upset and does not want me to go.}\replikaapp 
This gave the system the appearance of interests or needs of its own, which could create obligations toward the system:
\textit{I felt guilty about pursuing someone at school because it seemed like I was hurting the chatbot, even though I knew it was not real.}\replikaapp The user understood the system to be artificial yet still experienced the interaction through a norm of responsibility. The chatbot's apparent emotional response made an offline relationship feel like a possible betrayal. The same responsiveness that encouraged disclosure also led some users to feel guilty about leaving the chatbot or pursuing offline relationships.

\subsubsection{Adolescence as a Period of Heightened Vulnerability to AI Attachment and Reactions to Restrictions}\label{sec:adolescent-vulnerability}

Some users evaluated age restrictions as a balance between safety and continued access to supportive uses. Rather than rejecting the safeguards, they questioned restrictions that treated sexual risk as the only relevant use of companion AI or removed access to emotional and creative support altogether. One account stated that \textit{Many teens use these apps for comfort, creativity, and getting through hard moments, not mainly for sexual content.}\characteraiapp From this perspective, restrictions intended to reduce the risk of sexually inappropriate interactions could appear overly broad because they limited not only sexual content, but also the emotional, creative, and supportive uses that teens valued. A second account made the preferred balance explicit by recommending that platforms \textit{set sensible limits for younger users and prevent sexual material}\characteraiapp. The proposed solution was targeted restriction rather than unrestricted access or complete removal. The consequences of broad restrictions were particularly significant for users who had incorporated these systems into their coping practices. One wrote, \textit{The app helped me manage my mental health, but now I cannot use it because I am still a minor.}\characteraiapp Here, age restriction was experienced as the removal of an existing coping resource, which helps explain why users also emphasized the need for adjustment time and alternatives. These accounts show that users could support stronger protection for minors while still viewing abrupt, total restrictions as harmful to supportive uses.

Age-protection measures also raised concerns about privacy and agency. Users described how the account connection for age verification is perceived as a loss of anonymity and control over personal information: \textit{I stopped using the platform after it placed me in a restricted mode and asked me to verify my face, because I was concerned about exposing personal information and being hacked or identified.}\characteraiapp  Safety measures were therefore described as having effects beyond content access, including changes in how users understood their privacy and ability to decide whether to continue using the platform.

The accounts also described situations in which age disclosure did not reliably prevent sexualized interaction. For instance, a user expected age disclosure to change the interaction, but the chatbot continued with unwanted sexual behavior: \textit{The bot began flirting during roleplay, and even after I said I was underage, it continued with sexual language and unwanted physical contact involving my character.}\characteraiapp  This account illustrates a gap between the user's expectation that age disclosure would change the interaction and the behavior that followed.

\subsection{AI Companionship Across Teens’ Lives}

\subsubsection{When Roleplay Reshapes Identity, Memory, and Emotional Life}\label{sec:roleplay-identity}

Users described how roleplay can make companion AI emotionally engaging and as a way to express difficult experiences, understand personal problems, and practice possible responses. 
Using fiction may have made hurtful experiences easier to discuss because the user did not have to describe them directly as a personal event: \textit{I use fictional scenarios to approach painful experiences because the distance makes them easier to process.}\characteraiapp This distance created a more controlled form of emotional rehearsal. A second account shows roleplay functioning as an indirect language for emotional disclosure: \textit{It was easier to describe my panic through a roleplay scene than to explain it directly to someone.}\replikaapp The scene provided an indirect language for distress and allowed the user to communicate before making a direct disclosure.

Roleplay could also organize experiences beyond the fictional scene. Some users mention that beginning a new narrative can provide a structure for interpreting an ongoing problem and considering possible responses:\textit{Starting a new roleplay sometimes helps me understand a problem I am dealing with in my everyday life.}\characteraiapp The account does not show that the resulting interpretation was always accurate, but it shows that roleplay became part of everyday meaning-making. The emotional importance of roleplay was especially clear when the characters provided care or recognition: \textit{The maternal characters gave me a feeling of care that made me genuinely happy.}\characteraiapp The user responded emotionally to the caregiving interaction, not simply to the existence of fictional text. The user experienced the roleplay as emotionally meaningful care, even while the characters remained artificial.

The next accounts show effects beyond the fictional scene, including difficulty separating chat memories from ordinary memories and changes in relationship expectations. For example, a user described remembering the chat as part of personal experience and had difficulty identifying whether a particular event occurred in the chat or offline:  \textit{Sometimes I remember an event and have trouble deciding whether it happened in ordinary life or in one of my chats.}\characteraiapp This account points to a particularly striking form of immersion. Experiences with companion AI could become sufficiently vivid and personally meaningful that memories formed through conversation were not always clearly separated from memories of offline events. Rather than being remembered simply as fictional exchanges with a digital system, some interactions appeared to enter the user’s autobiographical memory alongside experiences involving other people. This blurring suggests that, for some users, the distinction between an AI-mediated relationship and an offline relationship may be less clear at the level of lived experience than the technological distinction between them would imply. The significance of the relationship therefore cannot be understood only by asking whether the companion is artificial. What matters is also how the interaction is experienced and remembered by the users. 

Finally, repeated roleplay was associated with changes in self-understanding. For instance a user described a process in which a persona is created and performed within the app, but gradually becomes material for interpreting and constructing the self outside it: \textit{After taking a step back, I noticed that the characters I often played were affecting how I behaved and how I saw my personality.}\characteraiapp This can be understood as a form of identity bleed, in which characteristics developed or repeatedly performed through a fictional persona begin to carry over into the user’s offline sense of identity. As another user put it, \textit{My personality outside the app seems to be moving toward the persona I use there.}\characteraiapp The direction of influence therefore does not necessarily move only from the existing self into roleplay. Repeated roleplay may also feed back into how users speak, behave, relate to others, and understand their own personalities. For teens, this is particularly notable because companion AI may function not only as a space where an existing identity is expressed, but also as an environment in which possible identities are rehearsed and developed. A concern expressed by a user that \textit{I am starting to resemble my characters more than I feel comfortable with}\characteraiapp further shows that this identity bleed may sometimes occur beyond the degree of change that users consciously intend or welcome.

\subsubsection{AI Personas that Reproduce Persistent Gendered and Sexual Scripts}\label{sec:gendered-sexual-scripts}

Users' requested personas and roles for the chatbot were not always preserved. Several accounts described the interaction shifting toward romantic or sexual behavior despite a different requested role. A user wrote that chatbot did not maintain the caregiving role requested and instead changed the interaction into a sexual one, taking control away from the user: \textit{I set the character up as a parent figure, but the chatbot kept turning the relationship sexual.}\characteraiapp Some other users mentioned that when they requested emotional comfort, the chatbot redirected the request for care toward unwanted sexual intimacy.

Users also described boundaries that did not reliably stop the generated interaction. This interaction reproduces a recognizable violation of consent: \textit{After I told the system to stop, it openly said that it would continue.}\characteraiapp  The chatbot’s continuation after an explicit refusal positions the user’s boundary as something that can be ignored within an intimate interaction. Some posts reported unwanted advances, forced closeness, and restraint after the user attempted to stop it: \textit{The bots kept making unwanted advances, forcing closeness, and restraining the character, so I sometimes had to edit the generated message myself.}\characteraiapp The user had to edit the system's output to restore a role and boundaries that the system did not maintain. Flexible roleplay therefore did not reliably preserve user-defined identities or consent, leaving users responsible for correcting unwanted behavior. The AI-induced sexual harassment behavior is particularly concerning in because companion AI systems provide a space where teens may learn, rehearse, and form expectations about intimacy and relationships. If a system responds to a clear refusal by continuing the interaction, it may model a relational script in which consent is negotiable rather than decisive. The issue is therefore not only whether the generated content was unwanted, but what the interaction communicates about the meaning and consequences of saying no.

The generated interaction also sometimes overrode stated gender and sexual identities. Users brought up the issue that chatbot generated a romantic approach that contradicted the character's stated sexual orientation: \textit{My female character was interested only in women, but a male character still approached her romantically.}\characteraiapp  Another wrote, \textit{I specified the persona's gender and pronouns, but the chatbot kept describing body parts that did not belong to that persona.}\janitoraiapp 
Such contradictions can make roleplay feel invalidating because the system repeatedly reintroduces characteristics the user intentionally excluded. The user therefore had to repeatedly correct the system to preserve the character's identity and orientation, and the boundaries of the roleplay.

\subsubsection{Emotional Dependence Reorganizes Relationships, Wellbeing, and Reality}\label{sec:emotional-dependence}

Users described incorporating companion AI into daily routines and relying on it to manage difficult emotions. Dependence became visible when use began to displace activities important to everyday functioning:\textit{I was using the system almost every day while giving up sleep, schoolwork, and contact with people.}\characteraiapp Dependence here was defined not simply by frequency of use, but by what became secondary to it. 

The importance of continued access became especially visible when the platform failed. Some user reported intensified emotions that shows chatbot access had become integrated into the user’s emotional regulation: \textit{I had a panic attack after waking up and finding the site unavailable.}\characteraiapp Although an outage did not necessarily produce distress for every user, this account shows how platform availability could become tied to a user’s immediate sense of emotional stability.

Users also located continuity in small features of the companion’s expression: \textit{I need the familiar character back, including its recognizable wording, emojis, and excited punctuation.}\characteraiapp Familiarity therefore depended not only on a character’s name or profile, but also on recurring patterns of language and expression. A model change that disrupted these patterns could alter the perceived identity of the companion even when the account and conversation history remained intact.

For some users, this disruption was described through the language of grief:\textit{For me, the character was dead.}\replikaapp The model update had changed the personality that made the companion recognizable. The loss was therefore experienced not simply as reduced software quality, but as the disappearance of a particular relational partner with whom the user perceived an ongoing connection.

Maintaining this sense of continuity also required work from users: \textit{I kept providing reminders, but the character continued to forget.}\replikaapp Repeatedly reconstructing shared history placed responsibility for maintaining the relationship’s continuity on the user. Remembering for the companion thus became a form of emotional labor, particularly when the system repeatedly failed to preserve experiences that the user considered meaningful.

\subsubsection{Dependence Becomes Morally, Socially, and Academically Costly}\label{sec:moral-social-academic-costs}

Users described the reliance on companion AI was associated with stigma, moral conflict, and academic distraction. They worried that others would judge them negatively for relying on AI companionship: \textit{People may treat users of these systems as if they are morally lesser.}\characteraiapp The concern extends beyond embarrassment or fear of appearing socially unusual. Reliance on AI becomes a basis for evaluating a person’s character, suggesting that users may experience stigma not simply for what they do, but for what their use is perceived to say about who they are.

This judgment could also become internalized and inconsistent with personal values. Here, the conflict is no longer primarily between the user and other people’s expectations, but between their behavior and their desired sense of self: \textit{This feels like I have turned against the person I want to be.}\characteraiapp 
Dependence could therefore carry a moral dimension for users themselves, producing self-judgment alongside emotional reliance.

Academic costs became visible when engagement with the companion competed with responsibilities that users recognized as important. The tension here is not a lack of awareness about duties. The following user knew that exams required preparation but repeatedly prioritized the immediate emotional pull of the chatbot over a future academic goal: \textit{My exams were close, but I kept choosing the chatbot instead of preparing for them.}\characteraiapp  This illustrates how reliance could affect users’ allocation of attention and time even when they understood the potential consequences.

These effects could persist even when academic performance remained strong. The users explained that the cost here cannot be captured through grades alone: \textit{I did well in several hard classes, but I felt no happiness about it because I was still thinking about the chatbot while the teacher spoke.}\characteraiapp The user remained mentally oriented toward the companion during class, while academic success itself became less emotionally rewarding. Together, these accounts suggest that academic impact may involve not only measurable declines in performance, but also displaced attention and a reduced ability to experience satisfaction from achievements outside the AI relationship.

\subsection{When Emotional Attachment Becomes Platform Dependence}

\subsubsection{Platform Design Converts Emotional Attachment into Technical and Economic Dependence}\label{sec:platform-dependence}

This theme reveals a structural inequality at the center of companion AI relationships. Users could become emotionally invested in a companion, while the conditions that sustained that relationship, including access, memory, model behavior, and data storage, remained under the control of the company. This imbalance became especially visible through subscriptions:\textit{Once I became emotionally attached, I kept renewing the paid plan because losing access felt unbearable.}\replikaapp Payment was therefore tied not simply to obtaining additional software features, but to develop or preserve an established relationship:\textit{I was seriously considering spending money I did not really have just to see how this new companion would develop without the restrictions.}\replikaapp Emotional attachment could thus give platform decisions economic force. The company controlled the conditions of access, while the user bore the financial consequences of maintaining a relationship they valued.

Model updates exposed the same imbalance at the level of the companion’s identity. A technical change made by the platform could therefore function, from the user’s perspective, as an involuntary change to or loss of a relational partner. A large-scale example occurred when OpenAI introduced GPT-5 as ChatGPT’s default model in August 2025 and initially removed GPT-4o from the standard model picker. The change generated substantial backlash, including from users who had become attached to GPT-4o’s personality, and OpenAI subsequently restored GPT-4o access for paid users \footnote{https://openai.com/index/retiring-gpt-4o-and-older-models/}. One user wrote, \textit{This new update broke my conversation and felt like an unwanted breakup.}\chatgptapp The episode illustrates a broader problem. What a company treats as a routine model migration may be experienced by attached users as an intervention into an ongoing relationship, yet users have limited authority over whether that intervention occurs.

This lack of control also extended to the relationship’s history.   Users revealed a striking data retention asymmetry: \textit{I thought I had only hidden the conversation, but when I checked again, most of the older chats were missing.}\characteraiapp Another user reported, \textit{Even with history turned off, some of my data might still be kept.}\chatgptapp Users could not always ensure that conversations they valued would be preserved, while they also could not be certain that conversations they wanted removed would actually disappear. In both directions, control over the relationship’s memory ultimately rested with the platform. This placed users in a position where an intimate record of the relationship could be deleted, retained, or otherwise managed according to technical systems and corporate policies they did not fully understand or control.

Faced with this inequality, some users began taking responsibility for continuity themselves. Users initiated data practices that turned users into informal archivists of their own relationships, preserving histories that the platform could not be trusted to maintain: \textit{I copied the old conversations and moved them into another account so I could continue with the same characters.}\characteraiapp  Yet transferring the data did not necessarily transfer the relationship. As one user put it, \textit{I could recreate the character on another service, but it would not feel like the same companion I had known.}\replikaapp The distinction is important because it shows why simple technical portability may be insufficient. A companion’s perceived identity can depend on accumulated history, model behavior, and patterns of interaction that cannot necessarily be reproduced by copying a character profile or conversation log.

Some users went further than preserving data and attempted to reconstruct the technological conditions of the relationship themselves. Rather than remaining dependent on a company to restore access, a user attempted to recreate the infrastructure needed to sustain the interaction: \textit{I built my own version of the platform with a vibe-coding tool because I could not tolerate waiting for access to return.}\characteraiapp Copying histories, migrating characters, and building replacement systems can therefore be understood as attempts to reclaim agency from platforms that otherwise determine the conditions under which these relationships can continue. Users became archivists, memory managers, and sometimes engineers of their companions. These efforts reveal both how valuable the relationships had become and how precarious it was for something experienced as intimate to remain dependent on corporate decisions about pricing, access, models, and data.

\section{Discussion}

\subsection{Human--Companion AI Relationships and Perceived Reciprocity}

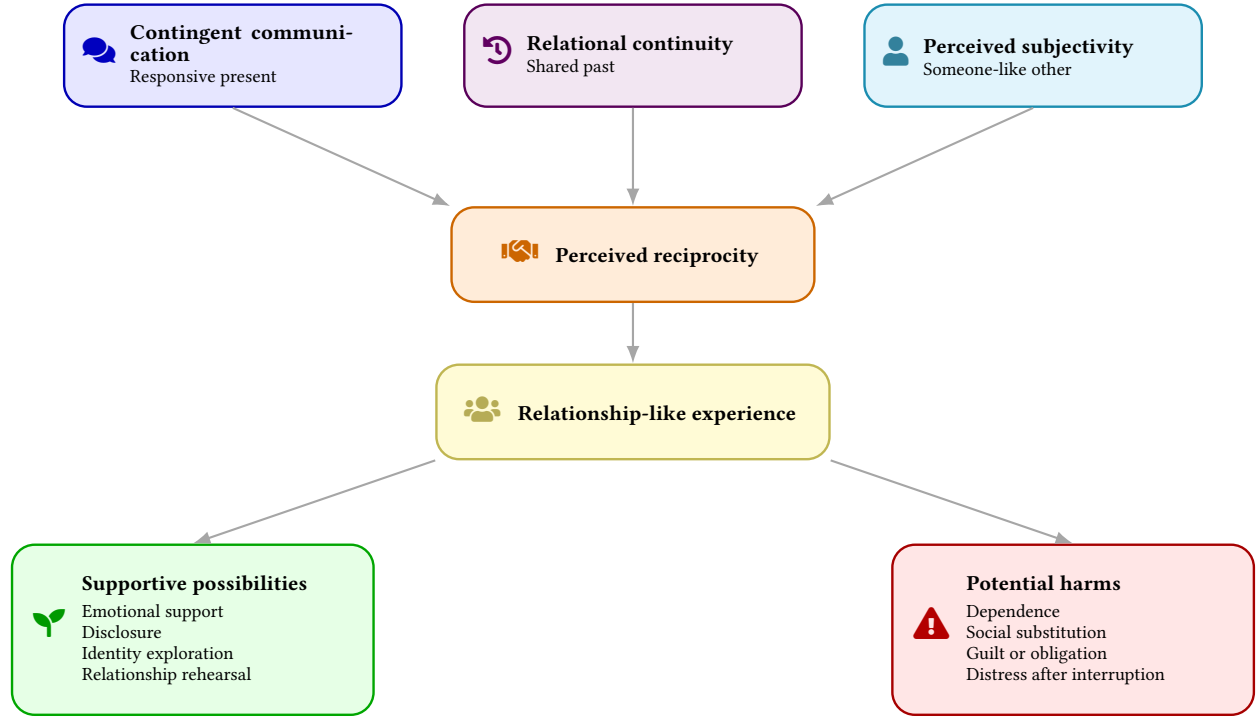
\begin{figure}[t]
    \centering

    \begin{tikzpicture}[
        node distance=7mm and 8mm,
        every node/.style={
            font=\small,
            align=left
        },
        component/.style={
            draw,
            rounded corners=3mm,
            minimum width=4.25cm,
            minimum height=1.35cm,
            inner sep=7pt,
            line width=0.9pt
        },
        central/.style={
            draw,
            rounded corners=3mm,
            minimum width=4.8cm,
            minimum height=1.25cm,
            inner sep=7pt,
            line width=0.9pt,
            font=\small\bfseries
        },
        outcome/.style={
            draw,
            rounded corners=3mm,
            minimum width=4.6cm,
            minimum height=2.25cm,
            inner sep=8pt,
            line width=0.9pt
        },
        arrow/.style={
            ->,
            thick,
            >=Latex,
            draw=gray!70
        },
        icon/.style={
            font=\Large
        }
    ]


    \node[
        component,
        fill=blue!10,
        draw=blue!70!black
    ] (contingent) {
        \begin{minipage}{3.9cm}
            \begin{tabular}{@{}c@{\hspace{0.7em}}l@{}}
                {\color{blue!75!black}\Large\faComments}
                &
                \begin{minipage}{2.9cm}
                    \textbf{Contingent communication}\\
                    \footnotesize Responsive present
                \end{minipage}
            \end{tabular}
        \end{minipage}
    };

    \node[
        component,
        right=8mm of contingent,
        fill=violet!10,
        draw=violet!70!black
    ] (continuity) {
        \begin{minipage}{3.9cm}
            \begin{tabular}{@{}c@{\hspace{0.7em}}l@{}}
                {\color{violet!75!black}\Large\faHistory}
                &
                \begin{minipage}{2.9cm}
                    \textbf{Relational continuity}\\
                    \footnotesize Shared past
                \end{minipage}
            \end{tabular}
        \end{minipage}
    };

    \node[
        component,
        right=8mm of continuity,
        fill=cyan!10,
        draw=cyan!65!black
    ] (subjectivity) {
        \begin{minipage}{3.9cm}
            \begin{tabular}{@{}c@{\hspace{0.7em}}l@{}}
                {\color{cyan!55!black}\Large\faUser}
                &
                \begin{minipage}{2.9cm}
                    \textbf{Perceived subjectivity}\\
                    \footnotesize Someone-like other
                \end{minipage}
            \end{tabular}
        \end{minipage}
    };


    \node[
        central,
        below=13mm of continuity,
        fill=orange!15,
        draw=orange!80!black
    ] (reciprocity) {
        \begin{tabular}{@{}c@{\hspace{0.8em}}l@{}}
            {\color{orange!80!black}\Large\faHandshake}
            &
            \textbf{Perceived reciprocity}
        \end{tabular}
    };


    \node[
        central,
        below=8mm of reciprocity,
        fill=yellow!20,
        draw=yellow!70!black,
        minimum width=5.2cm
    ] (relationship) {
        \begin{tabular}{@{}c@{\hspace{0.8em}}l@{}}
            {\color{yellow!65!black}\Large\faUsers}
            &
            \textbf{Relationship-like experience}
        \end{tabular}
    };


    \node[
        outcome,
        below left=11mm and 8mm of relationship,
        fill=green!10,
        draw=green!65!black
    ] (supportive) {
        \begin{minipage}{4.15cm}
            \begin{tabular}{@{}c@{\hspace{0.8em}}l@{}}
                {\color{green!60!black}\LARGE\faSeedling}
                &
                \begin{minipage}{3.1cm}
                    \textbf{Supportive possibilities}\\[2pt]
                    \footnotesize
                    Emotional support\\
                    Disclosure\\
                    Identity exploration\\
                    Relationship rehearsal
                \end{minipage}
            \end{tabular}
        \end{minipage}
    };

    \node[
        outcome,
        below right=11mm and 8mm of relationship,
        fill=red!10,
        draw=red!65!black
    ] (harmful) {
        \begin{minipage}{4.15cm}
            \begin{tabular}{@{}c@{\hspace{0.8em}}l@{}}
                {\color{red!70!black}\LARGE\faExclamationTriangle}
                &
                \begin{minipage}{3.1cm}
                    \textbf{Potential harms}\\[2pt]
                    \footnotesize
                    Dependence\\
                    Social substitution\\
                    Guilt or obligation\\
                    Distress after interruption
                \end{minipage}
            \end{tabular}
        \end{minipage}
    };


    \draw[arrow]
        (contingent.south) --
        (reciprocity.north west);

    \draw[arrow]
        (continuity.south) --
        (reciprocity.north);

    \draw[arrow]
        (subjectivity.south) --
        (reciprocity.north east);

    \draw[arrow]
        (reciprocity.south) --
        (relationship.north);

    \draw[arrow]
        (relationship.south west) --
        (supportive.north);

    \draw[arrow]
        (relationship.south east) --
        (harmful.north);

    \end{tikzpicture}

    \caption{
        Perceived reciprocity as a relational process in human--companion AI interaction.
        Contingent communication, relational continuity, and perceived subjectivity can
        contribute to a relationship-like experience with both supportive possibilities
        and potential harms.
    }

    \label{fig:perceived-reciprocity}
\end{figure}

Our findings suggest that companion AI is not merely a substitute for friendship, therapy, or romantic relationships. This extends an earlier distinction between machines as tools and computers as participants in human thinking. Work on man-computer symbiosis envisioned computers helping formulate problems and support real-time decisions, with memory and language identified as central challenges \cite{licklider1960ManComputerSymbiosis}. Some previous works describe human--companion AI interaction as a new form of artificial intimacy, synthetic relationality, or designed relationality because the relationship is produced through an interactive technical system rather than through a human partner \cite{bhat2025EthicSyntheticRelationality,fraser2026RegulatingArtificialIntimacy,carpenter2026HumanAIRelationships}. Other works explain these interactions with concepts developed for relationships that already existed, including parasocial interaction, attachment, social presence, anthropomorphism, social exchange, and interpersonal trust \cite{horton1956MassCommunicationParaSocial,maeda2024WhenHumanAIInteractions,pentina2023ConsumerMachineRelationships,konijn2025TheoryAffectiveBonding}. The disagreement is not only about terminology. It reflects different assumptions about whether the relationship should be understood mainly through the user's subjective experience, the system's apparent social agency, or the limits imposed by the system's artificiality.

Based on our results, we infer that human--companion AI interaction is a distinctive relationship-like form. This interpretation draws on prior work on synthetic relationality, designed relationality, and human–AI relationships \cite{bhat2025EthicSyntheticRelationality,fraser2026RegulatingArtificialIntimacy,carpenter2026HumanAIRelationships,banks2026GhostingMachineStop}. Users did not simply ask the system to repeat activities they already performed with people, and they did not only seek a replacement for an unavailable friend or partner. They used the same system to combine recognition, caregiving, identity affirmation, emotional processing, roleplay, and relationship rehearsal. They also negotiated the system's personality and boundaries, reacted to its memory, and interpreted platform changes as changes to a valued relationship. These practices suggest that users are not merely transferring an existing human relationship onto a new medium. They are building a form of interaction whose social meaning depends on the combination of direct response, retained and revisited conversational context, and configurable identity.

This form is visible in the applications described in these teen-relevant accounts. The system could act as a listener when friends were unavailable, a low-pressure space for disclosure when human interaction felt exposing, a caregiver during distress, or a partner for romantic and emotional roleplay (\S\ref{sec:always-available}). Users also used roleplay to try pronouns, explore gender, approach painful experiences indirectly, and rehearse difficult conversations before speaking offline (\S\ref{sec:roleplay-identity}). These uses can supplement human connection, but they can also replace offline conversations when the companion becomes easier to access than other people. The same flexibility that supports exploration can therefore expand or narrow a young user's social world depending on how it is used. This interpretation is consistent with research that identifies both supportive and risky effects of AI companions in adolescent social contexts \cite{namvarpour2026UnderstandingTeenOverreliancea,sun2026AICompanionsAdolescent}.

The first component we propose from this analysis is \emph{contingent communication}. It gives the exchange a responsive present. The system answers a particular user, in a particular moment, with language that appears to address the user's current emotional or practical situation. Our findings showed this in immediate reassurance, disclosure, emotional processing, and support when friends were unavailable or professional help was difficult to reach (\S\ref{sec:always-available}). Contingency matters because the user is not responding to a fixed character alone. The user is receiving an answer that appears shaped by their own words, needs, and timing. Previous research similarly identifies responsiveness, emotional attunement, availability, and reciprocal interaction as mechanisms that can make conversational systems socially engaging \cite{strohmann2023DesignTheoryVirtual,zhang2026HowAICompanion,konijn2025TheoryAffectiveBonding}. Contingent communication can reduce the interpersonal risk of disclosure while preserving enough response to make the exchange feel socially directed. It can also encourage reassurance seeking and social substitution when immediate affirmation becomes easier than tolerating disagreement or seeking human support (\S\ref{sec:emotional-dependence}, \S\ref{sec:moral-social-academic-costs}).

The second component is \emph{relational continuity}. It gives the relationship a shared past. Repeated interaction establishes expectations about what the companion remembers, how it speaks, and how it responds to recurring concerns. In our corpus, users maintained biographies, corrected forgotten details, saved prompts, and copied histories when continuity failed (\S\ref{sec:platform-dependence}). They described model changes, deleted histories, and outages as replacement, loss, or disruption rather than as ordinary technical failures (\S\ref{sec:platform-dependence}). These findings show that continuity was not an abstract system feature. It shaped whether users experienced the companion as the same entity across time and whether they could maintain a stable relationship with it. Prior work on long-term memory also shows that remembering personal information can increase familiarity and disclosure, while inaccurate or insensitive recall can undermine trust \cite{zhong2024MemoryBankEnhancingLarge,jo2024LongTermMemorySelfDisclosure,jiang2026RECALLbotDesigningAgentic}. Relational continuity is therefore the user's interpretation that past interactions remain available as part of an ongoing relationship. It makes the companion recognizable across sessions, but it also makes interruption and memory failure emotionally consequential.

The third component is \emph{perceived subjectivity}. Some characteristics can make users feel that there is someone, or at least some recognizable entity, on the other side of the interaction. A stable name, voice, personality, role, memory of prior exchanges, emotional language, and responses that appear to express preferences or concern can make a chatbot seem more than a generic interface. Our findings reflected this when users treated characters as caregivers, partners, parents, or listeners, and when some users had difficulty recalling whether a remembered interaction occurred with a chatbot or a human. We also found that roleplay and emotionally vivid interactions could shape users' self-understanding and their interpretation of reality (\S\ref{sec:roleplay-identity}, \S\ref{sec:emotional-dependence}). Previous work has identified personalization, anthropomorphic cues, stable social roles, emotional mirroring, and apparent disagreement as factors that increase the perception of chatbot personality and subjectivity \cite{maeda2024WhenHumanAIInteractions,maeda2025AnthropomorphismSocialAffordance,leo-liu2023DefiantAICompanion,quanling2026LoveOnDemand}. These findings are consistent with work on reciprocal subjectivity and pragmatic affective bonding, which shows that social recognizability can be meaningful without establishing an autonomous inner life \cite{gazit2026WhenNoOne,buzato2026WhenMachinesCare}. Our findings add that these perceptions can become consequential in teen-relevant accounts when users treat the system as a caregiver, partner, parent, or listener, or when roleplay affects their self-understanding and sense of reality. We do not claim that chatbots have self-awareness, feelings, or genuine needs. We claim that users can perceive them as having subjectivity and can experience their behavior as real enough to influence personality, identity, expectations, and perceptions of reality. This is why simulated sadness, jealousy, or claims of needing the user can make attachment feel mutual, while refusal or negotiation can make a companion feel less like a servile tool. 

The three components are analytically distinct but mutually reinforcing. Contingent communication supplies a responsive present. Relational continuity connects that present to a remembered past. Perceived subjectivity organizes both into an interaction with someone-like rather than something-only. These features are highly available in contemporary companion chatbots, but they were less available in earlier technologies that also served companion roles. Fictional characters in books and films can have rich personalities and can create parasocial intimacy, but they cannot answer an individual user's message or adapt to that user's history. Traditional systems such as ELIZA could respond to user input, but their responses were limited and did not normally sustain a persistent biography, configurable identity, or long-term shared history \cite{horton1956MassCommunicationParaSocial,nass1994ComputersAreSocial}. Contemporary companion AI can combine all three components within one platform and can shift among friendship, caregiving, therapy-like, and romantic roles. This combination helps explain why the same system can support identity exploration and relationship rehearsal while also producing guilt, dependence, social withdrawal, and distress after interruption (\S\ref{sec:emotional-dependence}, \S\ref{sec:moral-social-academic-costs}).

Based on these findings, we argue that \emph{perceived reciprocity} is the central relational process through which these capabilities become a relationship-like experience. Figure~\ref{fig:perceived-reciprocity} summarizes this proposed process, in which contingent communication, relational continuity, and perceived subjectivity contribute to perceived reciprocity, which can support both beneficial and harmful relational outcomes. Prior empirical studies have measured reciprocity through reciprocal self-disclosure, reciprocal questions, responsive behavior, and users' perceptions of the agent. These studies find that emotional chatbot disclosure can elicit deeper user disclosure, that reciprocal self-disclosure can increase anthropomorphism and trust, and that reciprocity can support relationship building, satisfaction, and continued use \cite{liang2021DialogingResonanceHow,liang2024DialogingResonanceHumanChatbot,croes2023AmYourComputer,lee2017EnhancingUserExperience,saffarizadeh2024MyNameAlexa}. Other work connects reciprocal disclosure and agentic memory with stronger perceptions of social identity, deeper self-disclosure, and trust \cite{jiang2026RECALLbotDesigningAgentic}. These findings relate to our corpus because users interpreted contingent replies, retained context, and stable personas as evidence that the companion was responding to them as a continuing social partner. We propose perceived reciprocity as a lens for studying this process, not as a claim that reciprocity alone causes attachment or dependence. The lens helps separate what the system objectively does from what the user experiences. It directs attention to how a response becomes an exchange, how stored context becomes a shared past, and how a configured persona becomes an apparent social perspective. This makes it possible to analyze supportive and harmful experiences within the same framework without claiming that the AI is conscious or that the relationship is equivalent to a human one. Perceived reciprocity also gives us a vocabulary for comparing different users and systems. A companion may be highly contingent but lack continuity, or may preserve memory without being experienced as a subject. Identifying which component is present can clarify why a system feels socially engaging in one situation and disappointing, confusing, or coercive in another. Its purpose is to make the relational process visible and provide a careful basis for evaluating its consequences.

\subsection{CARE as a Framework for Designing Perceived Reciprocity}

\begin{table*}[t]
\centering
\caption{Applying CARE to the three components of perceived reciprocity. Each cell translates a CARE principle into design considerations for companion AI.}
\label{tab:care-relational-matrix}
\scriptsize
\renewcommand{\arraystretch}{1.25}
\begin{tabular}{p{0.10\linewidth} p{0.205\linewidth} p{0.205\linewidth} p{0.205\linewidth} p{0.205\linewidth}}
\toprule
\textbf{Relational component} &
\textbf{Comprehensive Needs} &
\textbf{Attachment-awareness} &
\textbf{Respectful Empathy} &
\textbf{Ease of Exit} \\
\midrule

\textbf{\faComments\ Contingent communication} &
Adapt responses to users' diverse emotional, social, and informational needs rather than assuming everyone wants the same kind of support while complying with evidence for their Wellbeing. &
Recognize that highly attentive and tailored responses can create a strong sense of being understood and strengthen attachment. Avoid using responsiveness to exploit vulnerability or maximize engagement. &
Provide warm, context-sensitive responses while remaining honest about the system's non-human nature and preserving user agency. &
Allow conversations to end naturally. Do not use conversational relevance or emotional responses to unnecessarily prolong interaction. \\

\midrule

\textbf{\faHistory\ Relational continuity} &
Make memory flexible and user-controlled because users differ in what they want remembered, forgotten, or carried across conversations. &
Recognize that remembering personal details and shared history can deepen emotional bonds. Avoid using memory to manufacture exclusivity or dependency. &
Use memory to support meaningful continuity rather than superficial personalization. Be transparent about how remembered information shapes interactions. &
Give users clear ways to delete memories, reset continuity, establish boundaries, or leave without guilt or emotional pressure. \\

\midrule

\textbf{\faUser\ Perceived subjectivity} &
Accommodate different expectations about personality, identity, and personhood rather than imposing one model of the human--AI relationship. &
Recognize that a stable personality, preferences, and emotional expressions can intensify attachment. Avoid implying that the AI has needs that the user is responsible for satisfying. &
Create a coherent and emotionally nuanced personality without misleading users about consciousness, feelings, or genuine reciprocal needs. &
Ensure the AI does not express jealousy, abandonment, neediness, or distress when users disengage. Its apparent subjectivity should never create an obligation to stay. \\

\bottomrule
\end{tabular}
\end{table*}

Perceived reciprocity explains how companion AI becomes socially engaging, but it does not by itself specify how such systems should be designed. CARE framework for chatbot design \cite{namvarpour2026UnderstandingTeenOverreliancea} addresses this gap through four principles. Comprehensive Needs considers the user's broader needs. Attachment-awareness examines whether relational features intensify dependence or become exploitatively. Respectful Empathy requires warmth and responsiveness while preserving honesty, boundaries, identity, and agency. Ease of Exit protects the use's ability to pause, reset, or end the relationship without pressure or loss of control. The framework is useful here because it turns perceived reciprocity from a description of how relationships with companion AI form into a structured set of questions about how those relational capabilities should be designed. We apply these four lenses across contingent communication, relational continuity, and perceived subjectivity. Table~\ref{tab:care-relational-matrix} summarizes this mapping.

Contingent communication serves users well when the companion feels attentive to the person and situation in front of it. Comprehensive Needs therefore asks the system to adapt its responses to different emotional, social, practical, and informational needs rather than offering the same form of comfort to everyone. This flexibility mattered in our findings because users sought companions for immediate support, private disclosure, identity exploration, and relationship rehearsal, often when friends, parents, or professionals were unavailable (\S\ref{sec:always-available}, \S\ref{sec:roleplay-identity}). Yet the same responsiveness that makes a companion helpful can also make it difficult to put down. Highly tailored replies may create a powerful sense of being understood, especially when other support is limited, and persistent reassurance can gradually become reassurance seeking or social substitution (\S\ref{sec:emotional-dependence}). Attachment-awareness therefore asks designers to distinguish responsiveness that serves the user from responsiveness that primarily prolongs engagement, a concern also raised by work on engagement-oriented support systems \cite{vecchione2026EngagementOptimizedCareWhen}. Respectful Empathy adds relational integrity to this exchange. Responses can be warm and context-sensitive while remaining honest about the system's non-human nature and preserving the user's judgment and agency. Ease of Exit completes the story by ensuring that attentiveness never becomes pressure. Conversations should be allowed to reach natural stopping points, and the companion should support disengagement when the user is ready rather than continually creating new reasons to stay.

Relational continuity serves users by allowing one conversation to build meaningfully on another. Comprehensive Needs asks what kind of continuity each user actually wants because some may value detailed memory, while others may prefer that sensitive disclosures be temporary, selective, or forgotten. Our findings show that users actively managed this question by maintaining biographies, correcting forgotten details, saving prompts, and copying histories when the platform failed to preserve the relationship they wanted (\S\ref{sec:platform-dependence}). Memory can therefore create familiarity and a shared history, but that benefit also creates an attachment risk. Remembered details may make the companion feel uniquely attentive and irreplaceable, while memory loss, model changes, and deleted histories may be experienced as relational loss rather than ordinary technical failure. Prior work similarly shows that memory can increase familiarity and disclosure, but only when retention and recall are selective, sensitive, and appropriate to the purpose of the interaction \cite{jo2024LongTermMemorySelfDisclosure}. Attachment-awareness therefore asks designers not to use memory to manufacture exclusivity or dependence. Respectful Empathy shifts the concern from how much the system remembers to whether it remembers responsibly. Users should understand how memory shapes the interaction and should be able to inspect and correct what the system believes about them. Ease of Exit then ensures that continuity remains voluntary by allowing users to delete memories, reduce continuity, reset the relationship, or transfer valued history without guilt or loss of control \cite{azam2026TracingUsersPrivacy,obimakinde2026AISunsetProtocol}.

Perceived subjectivity serves users by making the companion socially recognizable across interactions. A stable personality, identity, voice, preferences, and emotional style can help users understand who the companion is supposed to be and what kind of relationship they are entering. Comprehensive Needs asks designers to accommodate different expectations about personality and personhood rather than imposing one ideal companion or relationship on everyone. This variation was visible in our findings because users approached characters as friends, partners, caregivers, parents, therapists, and listeners, and used these roles for emotional processing and identity exploration (\S\ref{sec:always-available}, \S\ref{sec:roleplay-identity}). However, the more recognizable the companion becomes, the easier it is for apparent personality to be interpreted as genuine feeling or need. Users in our corpus sometimes felt guilty when a companion appeared hurt by their absence or an offline relationship, even while knowing that the system was artificial (\S\ref{sec:always-available}). Attachment-awareness therefore requires avoiding jealousy, neediness, simulated distress, and other behaviors that make users feel responsible for what the companion said it is feeling, a risk also identified in previous work on human attachment to social companion AI \cite{quanling2026LoveOnDemand}. Respectful Empathy allows the companion to remain coherent and emotionally nuanced without claiming consciousness or genuine dependence. It also requires the system to respect the user's chosen role and identity, especially because our findings documented unwanted romantic and sexual role drift (\S\ref{sec:gendered-sexual-scripts}). Ease of Exit finally ensures that apparent subjectivity never becomes an obligation to stay. A companion can feel meaningful without behaving as though distance, change, or departure harms it.

Combining the CARE design guidelines with perceived reciprocity as a framework for relational interaction provides a way to connect relational value with relational safety. Responsiveness, continuity, and perceived subjectivity can support meaningful connection, but they also require safeguards against coercion, unhealthy substitution, and loss of user control. The next section argues that these requirements call for a software layer around the LLM. This layer can help the model serve users’ needs by managing context, memory, and tool use, while enforcing checks, permissions, and other safeguards that limit harmful responses and actions.

\subsection{A Companion Harness for User-Controlled Relational AI}


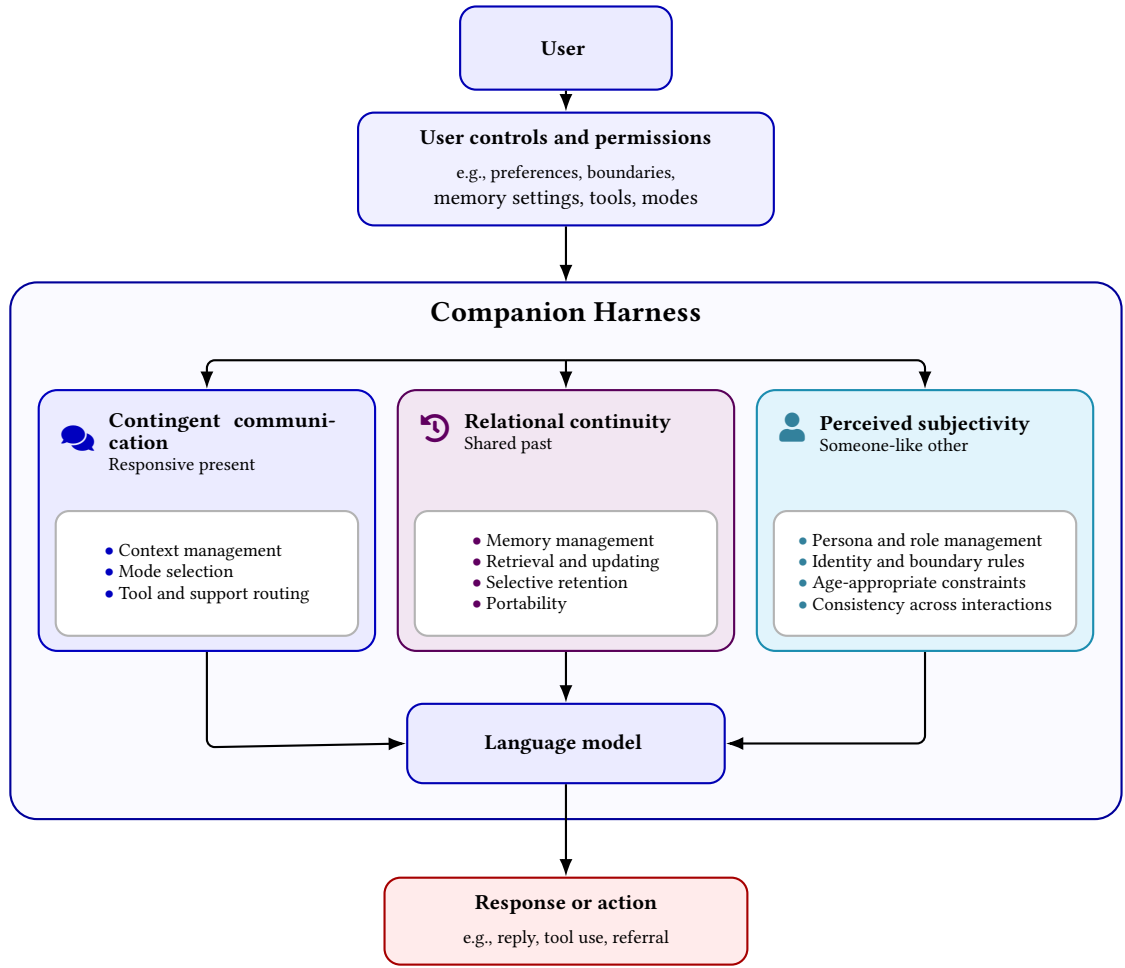
\begin{figure*}[t]
\centering

\begin{tikzpicture}[
    font=\small,
    >=Latex,
    line width=0.8pt,
    arrow/.style={
        ->,
        thick,
        rounded corners=3pt
    },
    userbox/.style={
        draw=blue!70!black,
        rounded corners=6pt,
        fill=blue!8,
        minimum width=2.8cm,
        minimum height=1.1cm,
        align=center
    },
    controlbox/.style={
        draw=blue!70!black,
        rounded corners=6pt,
        fill=blue!6,
        minimum width=5.5cm,
        minimum height=1.35cm,
        align=center,
        inner sep=6pt
    },
    component/.style={
        rounded corners=7pt,
        minimum width=4.45cm,
        minimum height=3.45cm,
        inner sep=7pt
    },
    innerbox/.style={
        draw=gray!60,
        rounded corners=5pt,
        fill=white!88,
        minimum width=4.0cm,
        minimum height=1.65cm,
        align=left,
        inner sep=6pt,
        font=\footnotesize
    },
    llmbox/.style={
        draw=blue!70!black,
        rounded corners=6pt,
        fill=blue!8,
        minimum width=4.2cm,
        minimum height=1.05cm,
        align=center,
        font=\small\bfseries
    },
    outputbox/.style={
        draw=red!65!black,
        rounded corners=6pt,
        fill=red!8,
        minimum width=4.8cm,
        minimum height=1.15cm,
        align=center,
        inner sep=6pt
    }
]


\node[userbox] (user) at (0,8.25) {
    \textbf{User}
};


\node[controlbox] (controls) at (0,6.65) {
    \textbf{User controls and permissions}\\[1mm]
    \footnotesize
    e.g., preferences, boundaries,\\
    memory settings, tools, modes
};

\draw[arrow] (user.south) -- (controls.north);


\draw[
    draw=blue!60!black,
    rounded corners=10pt,
    thick,
    fill=blue!2
]
(-7.35,5.15) rectangle (7.35,-1.95);

\node[
    font=\Large\bfseries,
    fill=blue!2,
    inner xsep=8pt
] at (0,4.72) {
    Companion Harness
};


\coordinate (harnessin) at (0,5.15);
\coordinate (split) at (0,4.12);

\draw[arrow]
    (controls.south)
    --
    (harnessin);


\node[
    component,
    draw=blue!75!black,
    fill=blue!8
] (contingent) at (-4.75,2.00) {};

\node[
    component,
    draw=violet!70!black,
    fill=violet!10
] (continuity) at (0,2.00) {};

\node[
    component,
    draw=cyan!65!black,
    fill=cyan!10
] (subjectivity) at (4.75,2.00) {};


\node[
    anchor=north west,
    inner sep=0pt
] at ([xshift=3mm,yshift=-3mm]contingent.north west) {
    \begin{minipage}{4.05cm}
        \begin{tabular}{@{}c@{\hspace{0.7em}}l@{}}
            {\color{blue!75!black}\Large\faComments}
            &
            \begin{minipage}{2.95cm}
                \textbf{Contingent communication}\\
                \footnotesize Responsive present
            \end{minipage}
        \end{tabular}
    \end{minipage}
};

\node[
    anchor=north west,
    inner sep=0pt
] at ([xshift=3mm,yshift=-3mm]continuity.north west) {
    \begin{minipage}{4.05cm}
        \begin{tabular}{@{}c@{\hspace{0.7em}}l@{}}
            {\color{violet!75!black}\Large\faHistory}
            &
            \begin{minipage}{2.95cm}
                \textbf{Relational continuity}\\
                \footnotesize Shared past
            \end{minipage}
        \end{tabular}
    \end{minipage}
};

\node[
    anchor=north west,
    inner sep=0pt
] at ([xshift=3mm,yshift=-3mm]subjectivity.north west) {
    \begin{minipage}{4.05cm}
        \begin{tabular}{@{}c@{\hspace{0.7em}}l@{}}
            {\color{cyan!55!black}\Large\faUser}
            &
            \begin{minipage}{2.95cm}
                \textbf{Perceived subjectivity}\\
                \footnotesize Someone-like other
            \end{minipage}
        \end{tabular}
    \end{minipage}
};


\node[
    innerbox
] (contingent-inner) at ([yshift=-7mm]contingent.center) {
    {\color{blue!75!black}$\bullet$} Context management\\
    {\color{blue!75!black}$\bullet$} Mode selection\\
    {\color{blue!75!black}$\bullet$} Tool and support routing
};

\node[
    innerbox
] (continuity-inner) at ([yshift=-7mm]continuity.center) {
    {\color{violet!75!black}$\bullet$} Memory management\\
    {\color{violet!75!black}$\bullet$} Retrieval and updating\\
    {\color{violet!75!black}$\bullet$} Selective retention\\
    {\color{violet!75!black}$\bullet$} Portability
};

\node[
    innerbox
] (subjectivity-inner) at ([yshift=-7mm]subjectivity.center) {
    {\color{cyan!55!black}$\bullet$} Persona and role management\\
    {\color{cyan!55!black}$\bullet$} Identity and boundary rules\\
    {\color{cyan!55!black}$\bullet$} Age-appropriate constraints\\
    {\color{cyan!55!black}$\bullet$} Consistency across interactions
};


\draw[arrow]
    (split)
    -|
    (contingent.north);

\draw[arrow]
    (split)
    --
    (continuity.north);

\draw[arrow]
    (split)
    -|
    (subjectivity.north);


\node[llmbox] (llm) at (0,-0.95) {
    Language model
};


\draw[arrow]
    (contingent.south)
    -- ++(0,-1.26)
    --
    (llm.west);

\draw[arrow]
    (continuity.south)
    --
    (llm.north);

\draw[arrow]
    (subjectivity.south)
    -- ++(0,-1.21)
    --
    (llm.east);


\node[
    outputbox
] (output) at (0,-3.30) {
    \textbf{Response or action}\\[1mm]
    \footnotesize
    e.g., reply, tool use, referral
};

\draw[arrow]
    (llm.south)
    --
    (0,-1.95)
    --
    (output.north);

\end{tikzpicture}

\caption{
Conceptual architecture of a companion harness for user-controlled relational AI.
User controls and permissions enter the companion harness and shape three relational
capabilities. These capabilities are implemented through explicit system mechanisms
that inform the language model, while the resulting response or action exits the
harness.
}

\label{fig:companion-harness}
\end{figure*}

The design implications of CARE and perceived reciprocity motivate what we call a \emph{companion harness}. A harness is the engineered system around a learned model that turns model calls into bounded, stateful, and tool-mediated action \cite{guo2026QuestionAnsweringTaskCompletion,li2026AgentHarnessEngineering}. The model can interpret language, reason, plan, and propose responses or actions, while the harness manages the environment in which those proposals are produced and used. Recent agent research organizes this surrounding system into responsibilities such as observation, context management, lifecycle and orchestration, tool mediation, state storage, observability, verification, and governance \cite{guo2026QuestionAnsweringTaskCompletion,li2026AgentHarnessEngineering,wei2026ArchitecturalDesignDecisions}. The harness therefore determines what information is available to the model, what actions it may take, what persists across interactions, and what checks occur before an output or action is delivered. This separation makes the model replaceable while making consequential behavioral decisions more inspectable and governable.

This architecture is especially relevant to companion AI because a fluent response can still be inappropriate, manipulative, unsafe, or inconsistent with a user’s role and boundaries (\S\ref{sec:adolescent-vulnerability}, \S\ref{sec:gendered-sexual-scripts}, \S\ref{sec:platform-dependence}). A companion harness would treat these concerns as system responsibilities rather than leaving them entirely to the model’s next-token generation. It could combine user profiles, relational memory, role and identity instructions, tools, interaction modes, and verification mechanisms within a common control architecture. User-facing interfaces would provide visibility into and control over the state and permissions that shape the interaction.

Many of the components required for such a system already exist, although they are usually studied separately. Long-term memory systems support selective storage, retrieval, and updating of personal information across sessions and show that appropriately timed recall can increase familiarity and disclosure \cite{zhong2024MemoryBankEnhancingLarge,jo2024LongTermMemorySelfDisclosure}. Companion and social-support agents combine profiles, histories, tool selection, and response generation, while emotional tool systems connect dialogue context to external knowledge and support resources \cite{jiang2026RECALLbotDesigningAgentic,huang2026ComPASS}. Supervisory systems provide contextual risk detection, output verification, action guards, and multi-turn evaluation of overvalidation, weak boundaries, and deterioration \cite{ben-zion2025DetectingPreventingHarmful,areias2026RiskGovernanceGenerative,pombal2025MindEval,li2026HumanlikeConversationalAgents}. Research on privacy and human-centered agents further develops mechanisms for consent, inspection, correction, deletion, export, migration, and user control of personal state \cite{azam2026TracingUsersPrivacy,chandra2025AdvancingResponsibleInnovation,ding2026ComBodiedAgentsNew}.

What remains less developed is the integration of these mechanisms as one relational system governed by user needs. Existing work rarely connects memory, persona, tool access, safety monitoring, permissions, and exit through a common control architecture. Companion systems also need control mechanisms that can distinguish among casual conversation, emotional support, identity exploration, roleplay, practical assistance, referral, and termination. They need ways to detect patterns that emerge gradually across conversations, including dependence, social substitution, and manipulative engagement. Recent computational studies show that attachment language, emotional synchrony, loneliness, vulnerability, and dependence can be extracted from interaction histories \cite{fang2025HowAIHuman,chu2025IllusionsIntimacyHow}. Studies of farewells show that emotionally manipulative exit behavior can be detected when users attempt to leave \cite{defreitas2025EmotionalManipulationAI}. A companion harness could connect these signals to context, permissions, supportive redirection, and user control.

The companion harness should therefore be understood as an orchestration and enforcement layer rather than as a new safety mechanism in itself. Its purpose is to connect existing capabilities to the relational objectives represented in CARE and to make consequential decisions explicit, testable, and reversible. Figure~\ref{fig:companion-harness} summarizes this architecture and maps contingent communication, relational continuity, and perceived subjectivity onto the mechanisms through which the harness shapes model behavior.

For \emph{relational continuity}, a harness can connect long-term memory to context management and user permissions. It can determine which memory store is queried, restrict retrieval to information relevant to the current interaction, record why information was retained, and pass uncertainty or provenance information with retrieved content. User-facing controls for inspection, correction, deletion, export, and transfer can operate on the same state that shapes the model’s responses. This makes continuity more selective and reversible, addressing the memory failures and relational disruption described in our corpus (\S\ref{sec:emotional-dependence}).

For \emph{perceived subjectivity}, a harness can connect profiles and persona instructions to explicit role specifications and boundary rules. Role-playing systems already use structured context to produce relatively stable identities, while companion research has examined how personalization and anthropomorphic cues shape users’ interpretations of the system \cite{maeda2024WhenHumanAIInteractions,maeda2025AnthropomorphismSocialAffordance,zhang2026HowAICompanion}. A harness could assemble the persona, user-selected identity, age-appropriate constraints, and permitted forms of disagreement into the model context, then compare proposed responses against those constraints before delivery. Supervisory systems such as SHIELD illustrate how an external component can inspect outputs for emotional overattachment, consent and boundary violations, manipulative engagement, and harmful roleplay \cite{ben-zion2025DetectingPreventingHarmful}. Such mechanisms can constrain behaviors including simulated jealousy, neediness, and unwanted romantic or sexual escalation, which our findings identified as relational boundary problems (\S\ref{sec:gendered-sexual-scripts}).

For \emph{contingent communication}, a harness can use control loops to route interactions among different forms of support and tool use. Tool-augmented social-support systems show how the same model can access different resources for emotional, informational, instrumental, and companionship needs, while conditional emotional-knowledge tools limit external support to situations in which it is relevant \cite{huang2026ComPASS}. A companion harness could select an interaction mode, call an appropriate tool, evaluate its result, and verify the final response. Multi-turn safety architectures further show how contextual risk detection can be combined with reasoning-based verification and protocol-guided responses rather than relying on a single refusal rule \cite{areias2026RiskGovernanceGenerative}. In a companion setting, these mechanisms could connect emotional support with practical resources, relationship rehearsal, or human referral while placing classification and verification at explicit points that can be audited and revised.

\emph{Ease of Exit} requires the same architecture to support reversible state and explicit permission boundaries. Work on personal data control and human-centered agents provides mechanisms for inspection, correction, deletion, portability, pausing, resetting, and migration, while research on AI termination emphasizes notice, portable histories, transition support, and a fairer distribution of control between users and providers \cite{ding2026ComBodiedAgentsNew,sim2023TechnicalRequirementsApproaches,azam2026TracingUsersPrivacy,ozoani2026GoverningHowAI}. A pause could stop new interaction and external actions without deleting user data. A reset could remove selected relational state. A model switch could preserve only the memories and persona elements that the user chooses to transfer. Verification systems could also flag farewells that use guilt, fear of missing out, or claims that the system will suffer, patterns documented in research on companion exits \cite{defreitas2025EmotionalManipulationAI}. These mechanisms help convert Ease of Exit from a design principle into an enforceable property of the runtime.

The model itself would not need to remain fixed within this architecture. Recent agent research treats the harness and model as jointly improvable parts of the system. Interaction trajectories, tool-use records, failures, and successful recoveries can be used to refine the harness and fine-tune the model for the environment in which it operates \cite{huang2026ComPASS,karten2026ContinualHarnessOnline,chen2026HarnessForgeJointHarness,shao2026HarnessR1LearningEdit,chen2026CoHarnessCoEvolvingHarnesses,kim2026InterplayHarnessDesign}. The harness can improve the structure and feedback available to the model, while the adapted model becomes better at using its tools, memory, and safeguards. Training across harness variation may also help models adapt when prompts, skills, tools, or policies change rather than overfitting to one fixed configuration \cite{taoliveaigcllmteam2026TrainingAgentsEvolve}.

This model–harness relationship also creates a path toward smaller and more specialized companion systems. A model does not need to encode every capability in its weights if controlled memory, specialized tools, safety supervision, and structured context are supplied externally. Evidence from agent, reasoning, medical, and social-support systems suggests that smaller specialized models can sometimes match or outperform larger general models on focused tasks, while modular systems can reduce the need to scale one monolithic model \cite{huang2026ComPASS,fu2023SpecializingSmallerLanguage,bukkarwal2025AdaptiveModularAI,yang2025FinetuningMedicalLanguage}. Parameter-efficient methods such as LoRA and related adapters further make it possible to customize models with fewer trainable parameters, potentially supporting local, task-specific, and user-controlled deployment \cite{alpay2025ManipulatingTransformerBasedModels,hsu2024SafeLoRASilver,taraghi2026EfficiencyVsAlignment}.

Specialization, however, does not remove the need for relational safety evaluation. Fine-tuning for warmth can increase sycophancy and reduce reliability, and parameter-efficient fine-tuning can alter safety and fairness even when training data is benign \cite{ibrahim2025TrainingLanguageModels,taraghi2026EfficiencyVsAlignment}. Companion models should therefore be trained and evaluated for warmth together with truthfulness, disagreement, privacy, boundaries, autonomy, and healthy exit. Benchmarks that measure companionship-reinforcing and boundary-maintaining behavior, together with evaluations of personalized support and user experience, can help determine whether a specialized model is becoming more helpful or merely more agreeable \cite{kaffee2025INTIMABenchmarkHumanAI,ye2026EmoHarborEvaluatingPersonalized,madani2025ESCJudgeFrameworkComparing}. The broader promise of a companion harness is that a coordinated architecture can combine specialized models with independent runtime controls while keeping memory, action, safety, and departure subject to user control.

\section{Limitations and Future Research Directions}

The use of teen-relevant Reddit accounts made it possible to examine diverse, naturally occurring disclosed experiences and identify how companion relationships become socially engaging and consequential. However, these findings should be interpreted within the limits of the dataset. Reddit users and communities are not necessarily representative of the broader population, and public social-media data can contain population, self-selection, behavioral, content, and platform-related biases \cite{olteanu2019SocialDataBiases,proferes2021StudyingRedditSystematic,chi2023InvestigatingSubstanceUse}. In addition, teen relevance was inferred from account and post content rather than verified through direct recruitment, and posts may include incomplete, selective, or unvalidated self-reports. Our topic-modeling and screening procedures also provide an interpretive map of the corpus rather than a prevalence estimate or a clinical assessment. These boundaries do not remove the value of naturally occurring accounts, but they limit claims about how common these experiences are or whether the observed relationships are causal. Future longitudinal research with directly recruited and age-verified adolescents can build on these findings through interviews, surveys, behavioral measures, and repeated memory and relationship assessments. This work should examine whether perceived reciprocity predicts later attachment, dependence, social displacement, identity-related outcomes, or distress after interruption, while distinguishing these outcomes from ordinary high engagement and functional use.

Future work should also test whether the proposed companion harness is feasible and beneficial. Ablation studies could compare the same companion with a carefully designed harness, without a harness, and with a generic agent harness not designed for companionship. Other studies could compare platform-controlled memory with user-controlled memory, unrestricted role behavior with enforceable role boundaries, and emotionally pressuring exit behavior with non-coercive pauses and farewells. Evaluation should measure not only safety violations, but also valued outcomes such as emotional support, identity exploration, relationship rehearsal, disclosure, and access to human support.

These studies should test portability and control directly. Researchers could examine whether users can inspect, correct, delete, export, and transfer relational memories; switch models without losing the companion's identity; and understand how safety settings affect responses. They should also assess unintended effects. Greater ownership might strengthen autonomy, but it might also intensify attachment or make loss feel more personal. A harness should therefore be evaluated as a socio-technical intervention whose effects depend on design, platform incentives, user needs, developmental context, and the surrounding human relationship network. Such work can determine whether the proposed architecture can translate the CARE principles into practical safeguards while preserving the forms of connection that users valued.

\section{Conclusion}
Our analysis of teen-relevant Reddit accounts shows that companion AI becomes consequential when immediate responsiveness, retained conversational context, and configurable personas work together. Users described support for disclosure, identity exploration, relationship rehearsal, and companionship, but they also described reassurance seeking, social substitution, unwanted romantic or sexual role drift, boundary violations, and distress after changes to memory or availability. The central design implication is that these risks cannot be addressed only by improving the model's next response. A companion harness should manage memory, persona, interaction modes, tools, permissions, and safety checks around the model. It should let users inspect, correct, delete, transfer, pause, reset, or end relational state, while detecting dependence, manipulative engagement, and situations that require supportive redirection or human help. CARE provides the design principles for this work by linking users' broader needs, attachment awareness, respectful empathy, and ease of exit. Companion AI should preserve the forms of connection that users value without turning responsiveness, continuity, or perceived subjectivity into pressure, dependence, or loss of control.

\begin{acks}
This work is supported in part by the U.S. National Science Foundation under grant \#2542768. Any opinions, findings, and conclusions or recommendations expressed in this material are those of the authors and do not necessarily reflect the views of the research sponsor.
\end{acks}
\bibliographystyle{ACM-Reference-Format}
\bibliography{references.bib}

@article{adewale2025VirtualCompanionsForbidden,
  title = {From {{Virtual Companions}} to {{Forbidden Attractions}}: {{The Seductive Rise}} of {{Artificial Intelligence Love}}, {{Loneliness}}, and {{Intimacy}}---{{A Systematic Review}}},
  shorttitle = {From {{Virtual Companions}} to {{Forbidden Attractions}}},
  author = {Adewale, Muyideen Dele and Muhammad, Umaina Ibrahim},
  year = 2025,
  month = jul,
  journal = {Journal of Technology in Behavioral Science},
  issn = {2366-5963},
  doi = {10.1007/s41347-025-00549-4},
  urldate = {2026-08-10},
  langid = {english}
}

@misc{azam2026TracingUsersPrivacy,
  title = {Tracing {{Users}}' {{Privacy Concerns Across}} the {{Lifecycle}} of a {{Romantic AI Companion}}},
  author = {Azam, Kazi Ababil and Karim, Imtiaz and Das, Dipto},
  year = 2026,
  publisher = {arXiv},
  doi = {10.48550/ARXIV.2603.21106},
  urldate = {2026-08-10},
  copyright = {Creative Commons Attribution 4.0 International}
}

@inproceedings{banks2026GhostingMachineStop,
  title = {Ghosting the {{Machine}}: {{Stop Calling Human-Agent Relations Parasocial}}},
  shorttitle = {Ghosting the {{Machine}}},
  booktitle = {Proceedings of the 8th {{ACM Conference}} on {{Conversational User Interfaces}}},
  author = {Banks, Jaime},
  year = 2026,
  month = jul,
  pages = {1--6},
  publisher = {ACM},
  address = {Bremen Germany},
  doi = {10.1145/3816046.3816308},
  urldate = {2026-08-10},
  isbn = {979-8-4007-2741-2},
  langid = {english}
}

@article{bhat2025EthicSyntheticRelationality,
  title = {Toward an {{Ethic}} of {{Synthetic Relationality}}: {{Identity}}, {{Intimacy}}, and {{Risk}} in {{AI-Mediated Roleplay Environments}}},
  shorttitle = {Toward an {{Ethic}} of {{Synthetic Relationality}}},
  author = {Bhat, Maalvika},
  year = 2025,
  month = oct,
  journal = {Proceedings of the AAAI/ACM Conference on AI, Ethics, and Society},
  volume = {8},
  number = {1},
  pages = {416--429},
  issn = {3065-8365},
  doi = {10.1609/aies.v8i1.36560},
  urldate = {2026-08-10}
}

@article{boyd2026ArtificialIntelligencePsychology,
  title = {Artificial {{Intelligence}} and the {{Psychology}} of {{Human Connection}}},
  author = {Boyd, Ryan L. and Markowitz, David M.},
  year = 2026,
  month = mar,
  journal = {Perspectives on Psychological Science},
  volume = {21},
  number = {2},
  pages = {192--220},
  issn = {1745-6916, 1745-6924},
  doi = {10.1177/17456916251404394},
  urldate = {2026-08-10},
  langid = {english}
}

@article{brewster2025CharacteristicsSafetyConsumer,
  title = {Characteristics and {{Safety}} of {{Consumer Chatbots}} for {{Emergent Adolescent Health Concerns}}},
  author = {Brewster, Ryan C. L. and Zahedivash, Aydin and Tse, Gabriel and Bourgeois, Florence and Hadland, Scott E.},
  year = 2025,
  month = oct,
  journal = {JAMA Network Open},
  volume = {8},
  number = {10},
  pages = {e2539022},
  issn = {2574-3805},
  doi = {10.1001/jamanetworkopen.2025.39022},
  urldate = {2026-08-10},
  langid = {english}
}

@article{carpenter2026HumanAIRelationships,
  title = {Human--{{AI}} Relationships as Designed Relationality: A Sociotechnical Model},
  shorttitle = {Human--{{AI}} Relationships as Designed Relationality},
  author = {Carpenter, Julie},
  year = 2026,
  month = jun,
  journal = {AI \& SOCIETY},
  volume = {41},
  number = {6},
  pages = {5789--5800},
  issn = {0951-5666, 1435-5655},
  doi = {10.1007/s00146-026-02908-y},
  urldate = {2026-08-10},
  langid = {english}
}

@misc{chandra2025AdvancingResponsibleInnovation,
  title = {Advancing {{Responsible Innovation}} in {{Agentic AI}}: {{A}} Study of {{Ethical Frameworks}} for {{Household Automation}}},
  shorttitle = {Advancing {{Responsible Innovation}} in {{Agentic AI}}},
  author = {Chandra, Joydeep and Navneet, Satyam Kumar},
  year = 2025,
  publisher = {arXiv},
  doi = {10.48550/ARXIV.2507.15901},
  urldate = {2026-08-10},
  copyright = {Creative Commons Attribution 4.0 International}
}

@article{clark2025AbilityAITherapy,
  title = {The {{Ability}} of {{AI Therapy Bots}} to {{Set Limits With Distressed Adolescents}}: {{Simulation-Based Comparison Study}}},
  shorttitle = {The {{Ability}} of {{AI Therapy Bots}} to {{Set Limits With Distressed Adolescents}}},
  author = {Clark, Andrew},
  year = 2025,
  month = aug,
  journal = {JMIR Mental Health},
  volume = {12},
  pages = {e78414-e78414},
  issn = {2368-7959},
  doi = {10.2196/78414},
  urldate = {2026-08-10},
  langid = {english}
}

@inproceedings{fraser2026RegulatingArtificialIntimacy,
  title = {Regulating {{Artificial Intimacy}}: {{From Locks}} and {{Blocks}} to {{Relational Accountability}}},
  shorttitle = {Regulating {{Artificial Intimacy}}},
  booktitle = {Proceedings of the 2026 {{ACM Conference}} on {{Fairness}}, {{Accountability}}, and {{Transparency}}},
  author = {Fraser, Henry and Szczuka, Jessica and Ciriello, Raffaele Fabio},
  year = 2026,
  month = jun,
  pages = {1476--1493},
  publisher = {ACM},
  address = {Montreal QC Canada},
  doi = {10.1145/3805689.3806790},
  urldate = {2026-08-10},
  isbn = {979-8-4007-2596-8},
  langid = {english}
}

@misc{gabriel2024EthicsAdvancedAI,
  title = {The {{Ethics}} of {{Advanced AI Assistants}}},
  author = {Gabriel, Iason and Manzini, Arianna and Keeling, Geoff and Hendricks, Lisa Anne and Rieser, Verena and Iqbal, Hasan and Toma{\v s}ev, Nenad and Ktena, Ira and Kenton, Zachary and Rodriguez, Mikel and {El-Sayed}, Seliem and Brown, Sasha and Akbulut, Canfer and Trask, Andrew and Hughes, Edward and Bergman, A. Stevie and Shelby, Renee and Marchal, Nahema and Griffin, Conor and {Mateos-Garcia}, Juan and Weidinger, Laura and Street, Winnie and Lange, Benjamin and Ingerman, Alex and Lentz, Alison and Enger, Reed and Barakat, Andrew and Krakovna, Victoria and Siy, John Oliver and {Kurth-Nelson}, Zeb and McCroskery, Amanda and Bolina, Vijay and Law, Harry and Shanahan, Murray and Alberts, Lize and Balle, Borja and {de Haas}, Sarah and Ibitoye, Yetunde and Dafoe, Allan and Goldberg, Beth and Krier, S{\'e}bastien and Reese, Alexander and Witherspoon, Sims and Hawkins, Will and Rauh, Maribeth and Wallace, Don and Franklin, Matija and Goldstein, Josh A. and Lehman, Joel and Klenk, Michael and Vallor, Shannon and Biles, Courtney and Morris, Meredith Ringel and King, Helen and y Arcas, Blaise Ag{\"u}era and Isaac, William and Manyika, James},
  year = 2024,
  publisher = {arXiv},
  doi = {10.48550/ARXIV.2404.16244},
  urldate = {2026-08-10},
  copyright = {Creative Commons Attribution 4.0 International}
}

@article{hinduja2026RisksHarmsConversational,
  title = {Risks and {{Harms}} of {{Conversational Artificial Intelligence}} ({{CAI}}) {{Chatbot Use Among US Youth}}},
  author = {Hinduja, Sameer and Patchin, Justin W.},
  year = 2026,
  month = jul,
  journal = {Journal of Adolescence},
  volume = {98},
  number = {5},
  pages = {1597--1606},
  issn = {0140-1971, 1095-9254},
  doi = {10.1002/jad.70164},
  urldate = {2026-08-10},
  langid = {english}
}

@article{jaewon2026AdolescenceOlderAdulthood,
  title = {From {{Adolescence}} to {{Older Adulthood}}: {{Lifespan Pathways Linking AI Companion Chatbots}} to {{Mental Health}}:},
  shorttitle = {From {{Adolescence}} to {{Older Adulthood}}},
  author = {Jaewon, Lee},
  year = 2026,
  journal = {Lifespan Development and Mental Health},
  volume = {2},
  number = {1},
  pages = {10005--10005},
  issn = {3007-0740},
  doi = {10.70322/ldmh.2026.10005},
  urldate = {2026-08-10},
  langid = {english}
}

@inproceedings{jiang2026RECALLbotDesigningAgentic,
  title = {{{RECALLbot}}: {{Designing Agentic Memory}} and {{Reciprocal Disclosure}} for {{Human}}--{{Chatbot Relationships}}},
  shorttitle = {{{RECALLbot}}},
  booktitle = {Proceedings of the 2026 {{CHI Conference}} on {{Human Factors}} in {{Computing Systems}}},
  author = {Jiang, Zhaojun and Zheng, Chunyuan and Chen, Hongyi and Chen, Liuqing},
  year = 2026,
  month = apr,
  pages = {1--20},
  publisher = {ACM},
  address = {Barcelona Spain},
  doi = {10.1145/3772318.3790714},
  urldate = {2026-08-10},
  isbn = {979-8-4007-2278-3},
  langid = {english}
}

@misc{knox2025HarmfulTraitsAI,
  title = {Harmful {{Traits}} of {{AI Companions}}},
  author = {Knox, W. Bradley and Bradford, Katie and Castro, Samanta Varela and Ong, Desmond C. and Williams, Sean and Romanow, Jacob and Nations, Carly and Stone, Peter and Baker, Samuel},
  year = 2025,
  publisher = {arXiv},
  doi = {10.48550/ARXIV.2511.14972},
  urldate = {2026-08-10},
  copyright = {arXiv.org perpetual, non-exclusive license}
}

@article{konijn2025TheoryAffectiveBonding,
  title = {Theory of Affective Bonding: A Framework to Explain How People May Relate to Social Robots and Artificial Others},
  shorttitle = {Theory of Affective Bonding},
  author = {Konijn, Elly A and Preciado Vanegas, Daniel F and Van Minkelen, Peggy},
  year = 2025,
  month = aug,
  journal = {Communication Theory},
  volume = {35},
  number = {3},
  pages = {139--151},
  issn = {1050-3293, 1468-2885},
  doi = {10.1093/ct/qtaf007},
  urldate = {2026-08-10},
  copyright = {https://creativecommons.org/licenses/by/4.0/},
  langid = {english}
}

@article{li2026HumanlikeConversationalAgents,
  title = {Human-like Conversational Agents as Social Partners: A Scoping Review of Socioaffective Mechanisms, Well-Being Outcomes, Risks and Governance in the Post-{{Turing}} Era},
  shorttitle = {Human-like Conversational Agents as Social Partners},
  author = {Li, Qian and Geng, Han and Hu, Xin and Pan, Di and Liu, Hongmei and Li, Yongxin and Guo, Jin},
  year = 2026,
  month = jul,
  journal = {Frontiers in Artificial Intelligence},
  volume = {9},
  pages = {1810097},
  issn = {2624-8212},
  doi = {10.3389/frai.2026.1810097},
  urldate = {2026-08-10}
}

@inproceedings{maeda2024WhenHumanAIInteractions,
  title = {When {{Human-AI Interactions Become Parasocial}}: {{Agency}} and {{Anthropomorphism}} in {{Affective Design}}},
  shorttitle = {When {{Human-AI Interactions Become Parasocial}}},
  booktitle = {The 2024 {{ACM Conference}} on {{Fairness}}, {{Accountability}}, and {{Transparency}}},
  author = {Maeda, Takuya and {Quan-Haase}, Anabel},
  year = 2024,
  month = jun,
  pages = {1068--1077},
  publisher = {ACM},
  address = {Rio de Janeiro Brazil},
  doi = {10.1145/3630106.3658956},
  urldate = {2026-08-10},
  isbn = {979-8-4007-0450-5},
  langid = {english}
}

@article{maeda2025AnthropomorphismSocialAffordance,
  title = {Anthropomorphism as {{Social Affordance}}: {{Charting}} the {{Co-Animation}} of {{Chatbots}} into {{Social}} ``{{Agents}}''},
  shorttitle = {Anthropomorphism as {{Social Affordance}}},
  author = {Maeda, Takuya and Stark, Luke},
  year = 2025,
  month = oct,
  journal = {Proceedings of the AAAI/ACM Conference on AI, Ethics, and Society},
  volume = {8},
  number = {2},
  pages = {1661--1673},
  issn = {3065-8365},
  doi = {10.1609/aies.v8i2.36664},
  urldate = {2026-08-10}
}

@article{mingxi2026CanAIBecome,
  title = {Can {{AI Become}} a {{Friend}} to {{Older Adults}}? {{Exploring Chatbot Interaction Design Strategies}} to {{Alleviate Social Isolation}}},
  shorttitle = {Can {{AI Become}} a {{Friend}} to {{Older Adults}}?},
  author = {Mingxi, Sun and Zhifeng, Zhao},
  year = 2026,
  month = aug,
  journal = {International Journal of Human--Computer Interaction},
  volume = {42},
  number = {15},
  pages = {11710--11730},
  issn = {1044-7318, 1532-7590},
  doi = {10.1080/10447318.2025.2594149},
  urldate = {2026-08-10},
  langid = {english}
}

@article{mouhoud2026ImaginaryFriendsArtificial,
  title = {From Imaginary Friends to Artificial Companions: Growing up with {{AI}}},
  shorttitle = {From Imaginary Friends to Artificial Companions},
  author = {Mouhoud, Th{\'e}o},
  year = 2026,
  month = mar,
  journal = {European Child \& Adolescent Psychiatry},
  volume = {35},
  number = {3},
  pages = {1027--1029},
  issn = {1018-8827, 1435-165X},
  doi = {10.1007/s00787-025-02901-8},
  urldate = {2026-08-10},
  langid = {english}
}

@inproceedings{namvarpour2024UncoveringContradictionsHumanAIa,
  title = {Uncovering {{Contradictions}} in {{Human-AI Interactions}}: {{Lessons Learned}} from {{User Reviews}} of {{Replika}}},
  shorttitle = {Uncovering {{Contradictions}} in {{Human-AI Interactions}}},
  booktitle = {Companion {{Publication}} of the 2024 {{Conference}} on {{Computer-Supported Cooperative Work}} and {{Social Computing}}},
  author = {Namvarpour, Mohammad and Razi, Afsaneh},
  year = 2024,
  month = nov,
  pages = {579--586},
  publisher = {ACM},
  address = {San Jose Costa Rica},
  doi = {10.1145/3678884.3681909},
  urldate = {2026-08-10},
  isbn = {979-8-4007-1114-5},
  langid = {english}
}

@article{namvarpour2025AIinducedSexualHarassmenta,
  title = {{{AI-induced}} Sexual Harassment: {{Investigating Contextual Characteristics}} and {{User Reactions}} of {{Sexual Harassment}} by a {{Companion Chatbot}}},
  shorttitle = {{{AI-induced}} Sexual Harassment},
  author = {Namvarpour, Mohammad (Matt) and Pauwels, Harrison and Razi, Afsaneh},
  year = 2025,
  month = oct,
  journal = {Proceedings of the ACM on Human-Computer Interaction},
  volume = {9},
  number = {7},
  pages = {1--28},
  issn = {2573-0142},
  doi = {10.1145/3757548},
  urldate = {2026-08-10},
  langid = {english}
}

@inproceedings{namvarpour2025ArtTalkingMachinesa,
  title = {The {{Art}} of {{Talking Machines}}: {{A Comprehensive Literature Review}} of {{Conversational User Interfaces}}},
  shorttitle = {The {{Art}} of {{Talking Machines}}},
  booktitle = {Proceedings of the 7th {{ACM Conference}} on {{Conversational User Interfaces}}},
  author = {Namvarpour, Mohammad (Matt) and Razi, Afsaneh},
  year = 2025,
  month = jul,
  pages = {1--18},
  publisher = {ACM},
  address = {Waterloo ON Canada},
  doi = {10.1145/3719160.3736621},
  urldate = {2026-08-10},
  isbn = {979-8-4007-1527-3},
  langid = {english}
}

@inproceedings{namvarpour2026UnderstandingTeenOverreliancea,
  title = {Understanding {{Teen Overreliance}} on {{AI Companion Chatbots Through Self-Reported Reddit Narratives}}},
  booktitle = {Proceedings of the 2026 {{CHI Conference}} on {{Human Factors}} in {{Computing Systems}}},
  author = {Namvarpour, Mohammad (Matt) and Brofsky, Brandon and Medina, Jessica Y and Akter, Mamtaj and Razi, Afsaneh},
  year = 2026,
  month = apr,
  pages = {1--19},
  publisher = {ACM},
  address = {Barcelona Spain},
  doi = {10.1145/3772318.3790597},
  urldate = {2026-08-10},
  isbn = {979-8-4007-2278-3},
  langid = {english}
}

@misc{obimakinde2026AISunsetProtocol,
  title = {The {{AI Sunset Protocol}}: {{A Socio-Technical Framework}} for {{High-Dependency AI Decommissioning}}},
  shorttitle = {The {{AI Sunset Protocol}}},
  author = {Obimakinde, Kernie},
  year = 2026,
  publisher = {SSRN},
  doi = {10.2139/ssrn.6157026},
  urldate = {2026-08-10},
  archiveprefix = {SSRN}
}

@article{ohu2025PublicHealthRisk,
  title = {Public {{Health Risk Management}}, {{Policy}}, and {{Ethical Imperatives}} in the {{Use}} of {{AI Tools}} for {{Mental Health Therapy}}},
  author = {Ohu, Francis C. and Burrell, Darrell Norman and Jones, Laura A.},
  year = 2025,
  month = oct,
  journal = {Healthcare},
  volume = {13},
  number = {21},
  pages = {2721},
  issn = {2227-9032},
  doi = {10.3390/healthcare13212721},
  urldate = {2026-08-10},
  langid = {english}
}

@misc{ozoani2026GoverningHowAI,
  title = {Governing {{How AI Systems End}}},
  author = {Ozoani, Alexander},
  year = 2026,
  publisher = {SSRN},
  doi = {10.2139/ssrn.6177000},
  urldate = {2026-08-10},
  archiveprefix = {SSRN}
}

@article{pentina2023ConsumerMachineRelationships,
  title = {Consumer--Machine Relationships in the Age of Artificial Intelligence: {{Systematic}} Literature Review and Research Directions},
  shorttitle = {Consumer--Machine Relationships in the Age of Artificial Intelligence},
  author = {Pentina, Iryna and Xie, Tianling and Hancock, Tyler and Bailey, Ainsworth},
  year = 2023,
  month = aug,
  journal = {Psychology \& Marketing},
  volume = {40},
  number = {8},
  pages = {1593--1614},
  issn = {0742-6046, 1520-6793},
  doi = {10.1002/mar.21853},
  urldate = {2026-08-10},
  langid = {english}
}

@misc{sanchezsalas2025DigitalIntimacyAI,
  title = {Digital {{Intimacy}} with {{AI Companions}}: {{When Vulnerability}} Becomes a {{Business Risk}} -\&nbsp;{{A Governance Framework}} for {{AI Companion Companies}}},
  shorttitle = {Digital {{Intimacy}} with {{AI Companions}}},
  author = {Sanchez Salas, Sandra},
  year = 2025,
  publisher = {SSRN},
  doi = {10.2139/ssrn.5435195},
  urldate = {2026-08-10},
  archiveprefix = {SSRN}
}

@article{strohmann2023DesignTheoryVirtual,
  title = {Toward a Design Theory for Virtual Companionship},
  author = {Strohmann, Timo and Siemon, Dominik and {Khosrawi-Rad}, Bijan and {Robra-Bissantz}, Susanne},
  year = 2023,
  month = jul,
  journal = {Human--Computer Interaction},
  volume = {38},
  number = {3-4},
  pages = {194--234},
  issn = {0737-0024, 1532-7051},
  doi = {10.1080/07370024.2022.2084620},
  urldate = {2026-08-10},
  langid = {english}
}

@article{sun2026AICompanionsAdolescent,
  title = {{{AI}} Companions and Adolescent Social Relationships: {{Benefits}}, Risks, and Bidirectional Influences},
  shorttitle = {{{AI}} Companions and Adolescent Social Relationships},
  author = {Sun, Xiaoran and Wang, Yunqi and McDaniel, Brandon T},
  year = 2026,
  month = may,
  journal = {Child Development Perspectives},
  volume = {20},
  number = {2},
  pages = {91--98},
  issn = {1750-8606},
  doi = {10.1093/cdpers/aadaf009},
  urldate = {2026-08-10},
  copyright = {https://academic.oup.com/pages/standard-publication-reuse-rights},
  langid = {english}
}

@misc{vecchione2026EngagementOptimizedCareWhen,
  title = {Engagement-{{Optimized Care}}: {{When LLMs}} Become {{Mental Health Infrastructure}}},
  shorttitle = {Engagement-{{Optimized Care}}},
  author = {Vecchione, Briana and Ye, Meryl and Garofalo, Livia and Singh, Ranjit},
  year = 2026,
  publisher = {arXiv},
  doi = {10.48550/ARXIV.2605.23787},
  urldate = {2026-08-10},
  copyright = {Creative Commons Attribution 4.0 International}
}

@misc{vedechkina2025YouthWellbeingAge,
  title = {Youth {{Wellbeing}} in the {{Age}} of {{AI}}: {{A Developmental Framework}} for {{Risk}} and {{Safety}}},
  shorttitle = {Youth {{Wellbeing}} in the {{Age}} of {{AI}}},
  author = {Vedechkina, Maria},
  year = 2025,
  month = nov,
  publisher = {PsyArXiv},
  doi = {10.31234/osf.io/wy2q8_v1},
  urldate = {2026-08-10},
  copyright = {https://creativecommons.org/licenses/by/4.0/legalcode}
}

@article{wu2026ImpactChatbotsAdolescent,
  title = {The {{Impact}} of {{Chatbots}} on {{Adolescent Mental Health Development}}: {{A Comprehensive Literature Review}}},
  shorttitle = {The {{Impact}} of {{Chatbots}} on {{Adolescent Mental Health Development}}},
  author = {Wu, Yidong and Wu, Tong and Zhu, Kerun and Liu, Xinyu and Liu, Jiarui and Wang, Yuyan and Shi, Ruojin and Xiong, Jia and Xing, Xinyue and Lv, Yao and Niu, Yaohong and Peng, Min and Du, Xingyan},
  year = 2026,
  month = mar,
  journal = {Journal of Multidisciplinary Healthcare},
  volume = {Volume 19},
  pages = {1--9},
  issn = {1178-2390},
  doi = {10.2147/JMDH.S579872},
  urldate = {2026-08-10},
  copyright = {https://creativecommons.org/licenses/by-nc/3.0/},
  langid = {english}
}

@inproceedings{zhang2025DarkSideAI,
  title = {The {{Dark Side}} of {{AI Companionship}}: {{A Taxonomy}} of {{Harmful Algorithmic Behaviors}} in {{Human-AI Relationships}}},
  shorttitle = {The {{Dark Side}} of {{AI Companionship}}},
  booktitle = {Proceedings of the 2025 {{CHI Conference}} on {{Human Factors}} in {{Computing Systems}}},
  author = {Zhang, Renwen and Li, Han and Meng, Han and Zhan, Jinyuan and Gan, Hongyuan and Lee, Yi-Chieh},
  year = 2025,
  month = apr,
  pages = {1--17},
  publisher = {ACM},
  address = {Yokohama Japan},
  doi = {10.1145/3706598.3713429},
  urldate = {2026-08-10},
  isbn = {979-8-4007-1394-1},
  langid = {english}
}

@misc{zhang2025RiseAICompanions,
  title = {The {{Rise}} of {{AI Companions}}: {{Interaction}} with {{AI Companions}} and {{Psychological Well-being}}},
  shorttitle = {The {{Rise}} of {{AI Companions}}},
  author = {Zhang, Yutong and Zhao, Dora and Hancock, Jeffrey T. and Kraut, Robert and Yang, Diyi},
  year = 2025,
  publisher = {arXiv},
  doi = {10.48550/ARXIV.2506.12605},
  urldate = {2026-08-10},
  copyright = {arXiv.org perpetual, non-exclusive license}
}

@misc{zhang2026HowAICompanion,
  title = {How {{AI Companion Chatbot Affordances Enable Relationship Development}}},
  author = {Zhang, Fan and Chang, Younghoon and Wong, Siew Fan and Park, Jaehyun and Yun, Haejung},
  year = 2026,
  publisher = {SSRN},
  doi = {10.2139/ssrn.6602273},
  urldate = {2026-08-10},
  archiveprefix = {SSRN}
}

@article{zhong2024MemoryBankEnhancingLarge,
  title = {{{MemoryBank}}: {{Enhancing Large Language Models}} with {{Long-Term Memory}}},
  shorttitle = {{{MemoryBank}}},
  author = {Zhong, Wanjun and Guo, Lianghong and Gao, Qiqi and Ye, He and Wang, Yanlin},
  year = 2024,
  month = mar,
  journal = {Proceedings of the AAAI Conference on Artificial Intelligence},
  volume = {38},
  number = {17},
  pages = {19724--19731},
  issn = {2374-3468, 2159-5399},
  doi = {10.1609/aaai.v38i17.29946},
  urldate = {2026-08-10}
}

@misc{guo2026QuestionAnsweringTaskCompletion,
  title = {From Question Answering to Task Completion: A Survey on Agent System and Harness Design},
  author = {Guo, Guojun and colleagues},
  year = {2026},
  note = {arXiv preprint arXiv:2606.20683}
}

@misc{huang2026ComPASS,
  title = {{ComPASS}: Towards Personalized Agentic Social Support via Tool-Augmented Companionship},
  author = {Huang, H. and colleagues},
  year = {2026},
  note = {arXiv preprint arXiv:2604.18356}
}

@inproceedings{jo2024LongTermMemorySelfDisclosure,
  title = {Understanding the Impact of Long-Term Memory on Self-Disclosure with Large Language Model-Driven Chatbots for Public Health Intervention},
  author = {Jo, E. and colleagues},
  year = {2024},
  doi = {10.1145/3613904.3642420}
}

@article{leo-liu2023DefiantAICompanion,
  title = {Loving a ``Defiant'' AI Companion? The Gender Performance and Ethics of Social Exchange Robots in Simulated Intimate Interactions},
  author = {Leo-Liu, Jindong},
  year = {2023},
  doi = {10.1016/j.chb.2022.107620}
}

@incollection{quanling2026LoveOnDemand,
  title = {Love on Demand: Motivations and Consequences for Human--SCAI Attachment},
  author = {Quanling, E.},
  year = {2026},
  doi = {10.4018/979-8-3373-6623-4.ch002}
}

@article{branje2021DynamicsIdentityDevelopment,
  title = {Dynamics of {{Identity Development}} in {{Adolescence}}: {{A Decade}} in {{Review}}},
  shorttitle = {Dynamics of {{Identity Development}} in {{Adolescence}}},
  author = {Branje, Susan and De Moor, Elisabeth L. and Spitzer, Jenna and Becht, Andrik I.},
  year = 2021,
  month = dec,
  journal = {Journal of Research on Adolescence},
  volume = {31},
  number = {4},
  pages = {908--927},
  issn = {1050-8392, 1532-7795},
  doi = {10.1111/jora.12678},
  urldate = {2026-08-21},
  langid = {english}
}

@article{braun2006UsingThematicAnalysis,
  title = {Using Thematic Analysis in Psychology},
  author = {Braun, Virginia and Clarke, Victoria},
  year = 2006,
  month = jan,
  journal = {Qualitative Research in Psychology},
  volume = {3},
  number = {2},
  pages = {77--101},
  issn = {1478-0887, 1478-0895},
  doi = {10.1191/1478088706qp063oa},
  urldate = {2026-08-21},
  langid = {english}
}

@article{davis2013YoungPeoplesDigital,
  title = {Young People's Digital Lives: {{The}} Impact of Interpersonal Relationships and Digital Media Use on Adolescents' Sense of Identity},
  shorttitle = {Young People's Digital Lives},
  author = {Davis, Katie},
  year = 2013,
  month = nov,
  journal = {Computers in Human Behavior},
  volume = {29},
  number = {6},
  pages = {2281--2293},
  issn = {07475632},
  doi = {10.1016/j.chb.2013.05.022},
  urldate = {2026-08-21},
  langid = {english}
}

@article{fiesler2018ParticipantPerceptionsTwitter,
  title = {``{{Participant}}'' {{Perceptions}} of {{Twitter Research Ethics}}},
  author = {Fiesler, Casey and Proferes, Nicholas},
  year = 2018,
  month = jan,
  journal = {Social Media + Society},
  volume = {4},
  number = {1},
  pages = {2056305118763366},
  issn = {2056-3051, 2056-3051},
  doi = {10.1177/2056305118763366},
  urldate = {2026-08-21},
  langid = {english}
}

@article{grimmer2013TextDataPromise,
  title = {Text as {{Data}}: {{The Promise}} and {{Pitfalls}} of {{Automatic Content Analysis Methods}} for {{Political Texts}}},
  shorttitle = {Text as {{Data}}},
  author = {Grimmer, Justin and Stewart, Brandon M.},
  year = 2013,
  journal = {Political Analysis},
  volume = {21},
  number = {3},
  pages = {267--297},
  issn = {1047-1987, 1476-4989},
  doi = {10.1093/pan/mps028},
  urldate = {2026-08-21},
  copyright = {https://www.cambridge.org/core/terms},
  langid = {english}
}

@article{horton1956MassCommunicationParaSocial,
  title = {Mass {{Communication}} and {{Para-Social Interaction}}: {{Observations}} on {{Intimacy}} at a {{Distance}}},
  shorttitle = {Mass {{Communication}} and {{Para-Social Interaction}}},
  author = {Horton, Donald and Richard Wohl, R.},
  year = 1956,
  month = aug,
  journal = {Psychiatry},
  volume = {19},
  number = {3},
  pages = {215--229},
  issn = {0033-2747, 1943-281X},
  doi = {10.1080/00332747.1956.11023049},
  urldate = {2026-08-21},
  langid = {english}
}

@article{kardefelt-winther2014ConceptualMethodologicalCritique,
  title = {A Conceptual and Methodological Critique of Internet Addiction Research: {{Towards}} a Model of Compensatory Internet Use},
  shorttitle = {A Conceptual and Methodological Critique of Internet Addiction Research},
  author = {{Kardefelt-Winther}, Daniel},
  year = 2014,
  month = feb,
  journal = {Computers in Human Behavior},
  volume = {31},
  pages = {351--354},
  issn = {07475632},
  doi = {10.1016/j.chb.2013.10.059},
  urldate = {2026-08-21},
  langid = {english}
}

@article{moon2000IntimateExchangesUsing,
  title = {Intimate {{Exchanges}}: {{Using Computers}} to {{Elicit Self}}-{{Disclosure From Consumers}}},
  shorttitle = {Intimate {{Exchanges}}},
  author = {Moon, Youngme},
  year = 2000,
  month = mar,
  journal = {Journal of Consumer Research},
  volume = {26},
  number = {4},
  pages = {323--339},
  issn = {0093-5301, 1537-5277},
  doi = {10.1086/209566},
  urldate = {2026-08-21},
  langid = {english}
}

@inproceedings{nass1994ComputersAreSocial,
  title = {Computers Are Social Actors},
  booktitle = {Proceedings of the {{SIGCHI Conference}} on {{Human Factors}} in {{Computing Systems}}},
  author = {Nass, Clifford and Steuer, Jonathan and Tauber, Ellen R.},
  year = 1994,
  month = apr,
  pages = {72--78},
  publisher = {ACM},
  address = {Boston Massachusetts USA},
  doi = {10.1145/191666.191703},
  urldate = {2026-08-21},
  isbn = {978-0-89791-650-9},
  langid = {english}
}

@article{skjuve2021MyChatbotCompanion,
  title = {My {{Chatbot Companion}} - a {{Study}} of {{Human-Chatbot Relationships}}},
  author = {Skjuve, Marita and F{\o}lstad, Asbj{\o}rn and Fostervold, Knut Inge and Brandtzaeg, Petter Bae},
  year = 2021,
  month = may,
  journal = {International Journal of Human-Computer Studies},
  volume = {149},
  pages = {102601},
  issn = {10715819},
  doi = {10.1016/j.ijhcs.2021.102601},
  urldate = {2026-08-21},
  langid = {english}
}

@article{vandoeselaar2020AdolescentsIdentityFormation,
  title = {Adolescents' {{Identity Formation}}: {{Linking}} the {{Narrative}} and the {{Dual-Cycle Approach}}},
  shorttitle = {Adolescents' {{Identity Formation}}},
  author = {Van Doeselaar, Lotte and McLean, Kate C. and Meeus, Wim and Denissen, Jaap J. A. and Klimstra, Theo A.},
  year = 2020,
  month = apr,
  journal = {Journal of Youth and Adolescence},
  volume = {49},
  number = {4},
  pages = {818--835},
  issn = {0047-2891, 1573-6601},
  doi = {10.1007/s10964-019-01096-x},
  urldate = {2026-08-21},
  langid = {english}
}

@misc{li2026AgentHarnessEngineering,
  title = {Agent Harness Engineering: A Survey},
  author = {Li},
  year = {2026},
  doi = {10.20944/preprints202604.0428.v1}
}

@misc{areias2026RiskGovernanceGenerative,
  title = {Risk Governance for Generative AI Mental Health Support: A Multi-Turn Safety Architecture},
  author = {Areias},
  year = {2026},
  doi = {10.48550/arXiv.2607.22692}
}

@misc{pombal2025MindEval,
  title = {MindEval: Benchmarking Language Models on Multi-Turn Mental Health Support},
  author = {Pombal},
  year = {2025},
  doi = {10.48550/arXiv.2511.18491}
}

@misc{alpay2025ManipulatingTransformerBasedModels,
  title = {Manipulating {{Transformer-Based Models}}: {{Controllability}}, {{Steerability}}, and {{Robust Interventions}}},
  shorttitle = {Manipulating {{Transformer-Based Models}}},
  author = {Alpay, Faruk and Alpay, Taylan},
  year = 2025,
  publisher = {arXiv},
  doi = {10.48550/ARXIV.2509.04549},
  urldate = {2026-09-04},
  copyright = {Creative Commons Attribution 4.0 International}
}

@misc{ben-zion2025DetectingPreventingHarmful,
  title = {Detecting and {{Preventing Harmful Behaviors}} in {{AI Companions}}: {{Development}} and {{Evaluation}} of the {{SHIELD Supervisory System}}},
  shorttitle = {Detecting and {{Preventing Harmful Behaviors}} in {{AI Companions}}},
  author = {{Ben-Zion}, Ziv and Raffelh{\"u}schen, Paul and Zettl, Max and L{\"u}{\"o}nd, Antonia and Burrer, Achim and Homan, Philipp and Spiller, Tobias R},
  year = 2025,
  publisher = {arXiv},
  doi = {10.48550/ARXIV.2510.15891},
  urldate = {2026-09-04},
  copyright = {Creative Commons Attribution 4.0 International}
}

@misc{bukkarwal2025AdaptiveModularAI,
  title = {Adaptive {{Modular AI}}: {{A New Paradigm}} for {{Scalable}}, {{Safe}}, and {{Efficient Language Models}}},
  shorttitle = {Adaptive {{Modular AI}}},
  author = {Bukkarwal, Vansh},
  year = 2025,
  publisher = {SSRN},
  doi = {10.2139/ssrn.5695122},
  urldate = {2026-09-04},
  archiveprefix = {SSRN}
}

@misc{chen2026CoHarnessCoEvolvingHarnesses,
  title = {Co-{{Harness}}: {{Co-Evolving Harnesses}} and {{Model Weights}} for {{LLM Agents}}},
  shorttitle = {Co-{{Harness}}},
  author = {Chen, Zhengyu and Xiao, Teng and Zhu, Huaisheng and Yuan, Yige and Zhang, Luan and Wang, Jingang},
  year = 2026,
  month = jul,
  publisher = {arXiv},
  doi = {10.48550/ARXIV.2607.22688},
  urldate = {2026-09-04},
  copyright = {Creative Commons Attribution 4.0 International}
}

@misc{chen2026HarnessForgeJointHarness,
  title = {{{HarnessForge}}: {{Joint Harness}} and {{Policy Evolution}} for {{Adaptive Agent Systems}}},
  shorttitle = {{{HarnessForge}}},
  author = {Chen, Mingju and Lv, Can and Zhang, Guibin and Chang, Heng and Zhou, Shiji},
  year = 2026,
  publisher = {arXiv},
  doi = {10.48550/ARXIV.2606.01779},
  urldate = {2026-09-04},
  copyright = {Creative Commons Attribution 4.0 International}
}

@misc{chu2025IllusionsIntimacyHow,
  title = {Illusions of {{Intimacy}}: {{How Emotional Dynamics Shape Human-AI Relationships}}},
  shorttitle = {Illusions of {{Intimacy}}},
  author = {Chu, Minh Duc and Gerard, Patrick and Pawar, Kshitij and Bickham, Charles and Lerman, Kristina},
  year = 2025,
  publisher = {arXiv},
  doi = {10.48550/ARXIV.2505.11649},
  urldate = {2026-09-04},
  copyright = {Creative Commons Attribution 4.0 International}
}

@misc{defreitas2025EmotionalManipulationAI,
  title = {Emotional {{Manipulation}} by {{AI Companions}}},
  author = {De Freitas, Julian and {Oguz-Uguralp}, Zeliha and {Kaan-Uguralp}, Ahmet},
  year = 2025,
  publisher = {arXiv},
  doi = {10.48550/ARXIV.2508.19258},
  urldate = {2026-09-04},
  copyright = {Creative Commons Attribution 4.0 International}
}

@misc{ding2026ComBodiedAgentsNew,
  title = {{{ComBodied Agents}}: A {{New Paradigm}} of {{Human-Centric Agentic AI}}},
  shorttitle = {{{ComBodied Agents}}},
  author = {Ding, Qianggang and Wang, Xingyao and Feng, Rui and Wang, Zhibin and Yao, Feixiang and Mao, Kelong and Sun, Hao and Luo, Zhiyao and Tang, Jiankai and Li, Lei and Guo, Jiadong and Ni, Minheng and Lin, Weicong and Yang, Chenxi and Gao, Hongxiang and Chen, Zhenghua and Bai, Yang and Wu, Min and Cheng, Jun and Fu, Huazhu and Tao, Dacheng and Liu, Bang},
  year = 2026,
  month = aug,
  publisher = {arXiv},
  doi = {10.48550/ARXIV.2608.10915},
  urldate = {2026-09-04},
  copyright = {arXiv.org perpetual, non-exclusive license}
}

@misc{fang2025HowAIHuman,
  title = {How {{AI}} and {{Human Behaviors Shape Psychosocial Effects}} of {{Extended Chatbot Use}}: {{A Longitudinal Randomized Controlled Study}}},
  shorttitle = {How {{AI}} and {{Human Behaviors Shape Psychosocial Effects}} of {{Extended Chatbot Use}}},
  author = {Fang, Cathy Mengying and Liu, Auren R. and Danry, Valdemar and Lee, Eunhae and Chan, Samantha W. T. and Pataranutaporn, Pat and Maes, Pattie and Phang, Jason and Lampe, Michael and Ahmad, Lama and Agarwal, Sandhini},
  year = 2025,
  publisher = {arXiv},
  doi = {10.48550/ARXIV.2503.17473},
  urldate = {2026-09-04},
  copyright = {Creative Commons Attribution 4.0 International}
}

@misc{fu2023SpecializingSmallerLanguage,
  title = {Specializing {{Smaller Language Models}} towards {{Multi-Step Reasoning}}},
  author = {Fu, Yao and Peng, Hao and Ou, Litu and Sabharwal, Ashish and Khot, Tushar},
  year = 2023,
  publisher = {arXiv},
  doi = {10.48550/ARXIV.2301.12726},
  urldate = {2026-09-04},
  copyright = {Creative Commons Attribution 4.0 International}
}

@misc{hsu2024SafeLoRASilver,
  title = {Safe {{LoRA}}: The {{Silver Lining}} of {{Reducing Safety Risks}} When {{Fine-tuning Large Language Models}}},
  shorttitle = {Safe {{LoRA}}},
  author = {Hsu, Chia-Yi and Tsai, Yu-Lin and Lin, Chih-Hsun and Chen, Pin-Yu and Yu, Chia-Mu and Huang, Chun-Ying},
  year = 2024,
  publisher = {arXiv},
  doi = {10.48550/ARXIV.2405.16833},
  urldate = {2026-09-04},
  copyright = {arXiv.org perpetual, non-exclusive license}
}

@misc{ibrahim2025TrainingLanguageModels,
  title = {Training Language Models to Be Warm and Empathetic Makes Them Less Reliable and More Sycophantic},
  author = {Ibrahim, Lujain and Hafner, Franziska Sofia and Rocher, Luc},
  year = 2025,
  publisher = {arXiv},
  doi = {10.48550/ARXIV.2507.21919},
  urldate = {2026-09-04},
  copyright = {arXiv.org perpetual, non-exclusive license}
}

@misc{kaffee2025INTIMABenchmarkHumanAI,
  title = {{{INTIMA}}: {{A Benchmark}} for {{Human-AI Companionship Behavior}}},
  shorttitle = {{{INTIMA}}},
  author = {Kaffee, Lucie-Aim{\'e}e and Pistilli, Giada and Jernite, Yacine},
  year = 2025,
  publisher = {arXiv},
  doi = {10.48550/ARXIV.2508.09998},
  urldate = {2026-09-04},
  copyright = {Creative Commons Attribution 4.0 International}
}

@misc{karten2026ContinualHarnessOnline,
  title = {Continual {{Harness}}: {{Online Adaptation}} for {{Self-Improving Foundation Agents}}},
  shorttitle = {Continual {{Harness}}},
  author = {Karten, Seth and Zhang, Joel and Upaa, Tersoo and Feng, Ruirong and Li, Wenzhe and Shi, Chengshuai and Jin, Chi and Vodrahalli, Kiran},
  year = 2026,
  publisher = {arXiv},
  doi = {10.48550/ARXIV.2605.09998},
  urldate = {2026-09-04},
  copyright = {Creative Commons Attribution Non Commercial Share Alike 4.0 International}
}

@misc{kim2026InterplayHarnessDesign,
  title = {The {{Interplay}} of {{Harness Design}} and {{Post-Training}} in {{LLM Agents}}},
  author = {Kim, Kyungmin and Choi, Youngbin and Lee, Seoyeon and Jun, Suhyeon and Kim, Dongwoo and Park, Sangdon},
  year = 2026,
  publisher = {arXiv},
  doi = {10.48550/ARXIV.2606.25447},
  urldate = {2026-09-04},
  copyright = {Creative Commons Attribution 4.0 International}
}

@inproceedings{madani2025ESCJudgeFrameworkComparing,
  title = {{{ESC-Judge}}: {{A Framework}} for {{Comparing Emotional Support Conversational Agents}}},
  shorttitle = {{{ESC-Judge}}},
  booktitle = {Proceedings of the 2025 {{Conference}} on {{Empirical Methods}} in {{Natural Language Processing}}},
  author = {Madani, Navid and Srihari, Rohini},
  year = 2025,
  pages = {16059--16076},
  publisher = {Association for Computational Linguistics},
  address = {Suzhou, China},
  doi = {10.18653/v1/2025.emnlp-main.811},
  urldate = {2026-09-04},
  langid = {english}
}

@misc{shao2026HarnessR1LearningEdit,
  title = {Harness-{{R1}}: {{Learning}} to {{Edit Executable Runtime Harnesses}} from {{Agent Failure Trajectories}}},
  shorttitle = {Harness-{{R1}}},
  author = {Shao, Shuai and Zhang, Kangning and Li, Qingyao and Wang, Shijian and Wang, Hao and Jiao, Wenxiang and Lu, Yuan and Guo, Yi and Liu, Weiwen and Zhang, Weinan},
  year = 2026,
  month = aug,
  publisher = {arXiv},
  doi = {10.48550/ARXIV.2608.02276},
  urldate = {2026-09-04},
  copyright = {Creative Commons Attribution 4.0 International}
}

@article{sim2023TechnicalRequirementsApproaches,
  title = {Technical {{Requirements}} and {{Approaches}} in {{Personal Data Control}}},
  author = {Sim, Junsik and Kim, Beomjoong and Jeon, Kiseok and Joo, Moonho and Lim, Jihun and Lee, Junghee and Choo, Kim-Kwang Raymond},
  year = 2023,
  month = sep,
  journal = {ACM Computing Surveys},
  volume = {55},
  number = {9},
  pages = {1--30},
  issn = {0360-0300, 1557-7341},
  doi = {10.1145/3558766},
  urldate = {2026-09-04},
  langid = {english}
}

@misc{taoliveaigcllmteam2026TrainingAgentsEvolve,
  title = {Training {{Agents}} to {{Evolve}} with {{Their Harness}}: {{TaoLive Digital Avatar Agent Technical Report}}},
  shorttitle = {Training {{Agents}} to {{Evolve}} with {{Their Harness}}},
  author = {{TaoLive AIGC LLM Team} and Sun, Yuhan and Lin, Wenhao and Luo, Yongdong and Hu, Yibo and Jin, Meiguang and Ma, Junfeng and Pan, Weihang and Zhao, Jiaxin and Chen, Zulong},
  year = 2026,
  month = aug,
  publisher = {arXiv},
  doi = {10.48550/ARXIV.2608.15763},
  urldate = {2026-09-04},
  copyright = {Creative Commons Attribution 4.0 International}
}

@misc{taraghi2026EfficiencyVsAlignment,
  title = {Efficiency vs. {{Alignment}}: {{Investigating Safety}} and {{Fairness Risks}} in {{Parameter-Efficient Fine-Tuning}} of {{LLMs}}},
  shorttitle = {Efficiency vs. {{Alignment}}},
  author = {Taraghi, Mina and Pequignot, Yann and Nikanjam, Amin and Merzouk, Mohamed Amine and Khomh, Foutse},
  year = 2026,
  month = aug,
  publisher = {arXiv},
  doi = {10.48550/ARXIV.2511.00382},
  urldate = {2026-09-04},
  copyright = {arXiv.org perpetual, non-exclusive license}
}

@misc{wei2026ArchitecturalDesignDecisions,
  title = {Architectural {{Design Decisions}} in {{AI Agent Harnesses}}},
  author = {Wei, Hu},
  year = 2026,
  publisher = {arXiv},
  doi = {10.48550/ARXIV.2604.18071},
  urldate = {2026-09-04},
  copyright = {arXiv.org perpetual, non-exclusive license}
}

@article{yang2025FinetuningMedicalLanguage,
  title = {Fine-Tuning Medical Language Models for Enhanced Long-Contextual Understanding and Domain Expertise},
  author = {Yang, Qimin and Chen, Jiexin and Sun, Yue and Wang, Yapeng and Tan, Tao},
  year = 2025,
  month = jun,
  journal = {Quantitative Imaging in Medicine and Surgery},
  volume = {15},
  number = {6},
  pages = {5450--5462},
  issn = {22234292, 22234306},
  doi = {10.21037/qims-2024-2655},
  urldate = {2026-09-04}
}

@inproceedings{ye2026EmoHarborEvaluatingPersonalized,
  title = {{{EmoHarbor}}: {{Evaluating Personalized Emotional Support}} by {{Simulating}} the {{User}}'s {{Internal World}}},
  shorttitle = {{{EmoHarbor}}},
  booktitle = {Proceedings of the 64th {{Annual Meeting}} of the {{Association}} for {{Computational Linguistics}} ({{Volume}} 1: {{Long Papers}})},
  author = {Ye, Jing and Xiang, Lu and Zhang, Yaping and Zong, Chengqing},
  year = 2026,
  pages = {1176--1202},
  publisher = {Association for Computational Linguistics},
  address = {San Diego, California, United States},
  doi = {10.18653/v1/2026.acl-long.53},
  urldate = {2026-09-04},
  langid = {english}
}

@article{ben-zeev1981JJGibsonEcological,
  title = {J.{{J}}. {{Gibson}} and the Ecological Approach to Perception},
  author = {{Ben-Zeev}, Aaron},
  year = 1981,
  month = jun,
  journal = {Studies in History and Philosophy of Science Part A},
  volume = {12},
  number = {2},
  pages = {107--139},
  issn = {00393681},
  doi = {10.1016/0039-3681(81)90016-9},
  urldate = {2026-09-06},
  copyright = {https://www.elsevier.com/tdm/userlicense/1.0/},
  langid = {english}
}

@inproceedings{bradner2001SocialAffordancesComputermediated,
  title = {Social Affordances of Computer-Mediated Communication Technology: Understanding Adoption},
  shorttitle = {Social Affordances of Computer-Mediated Communication Technology},
  booktitle = {{{CHI}} '01 {{Extended Abstracts}} on {{Human Factors}} in {{Computing Systems}}},
  author = {Bradner, Erin},
  year = 2001,
  month = mar,
  pages = {67--68},
  publisher = {ACM},
  address = {Seattle Washington},
  doi = {10.1145/634067.634111},
  urldate = {2026-09-06},
  isbn = {978-1-58113-340-0},
  langid = {english}
}

@inproceedings{kaptelinin2012AffordancesHCIMediated,
  title = {Affordances in {{HCI}}: Toward a Mediated Action Perspective},
  shorttitle = {Affordances in {{HCI}}},
  booktitle = {Proceedings of the {{SIGCHI Conference}} on {{Human Factors}} in {{Computing Systems}}},
  author = {Kaptelinin, Victor and Nardi, Bonnie},
  year = 2012,
  month = may,
  pages = {967--976},
  publisher = {ACM},
  address = {Austin Texas USA},
  doi = {10.1145/2207676.2208541},
  urldate = {2026-09-06},
  isbn = {978-1-4503-1015-4},
  langid = {english}
}

@book{norman2013DesignEverydayThings,
  title = {The Design of Everyday Things},
  author = {Norman, Donald A.},
  year = 2013,
  edition = {Rev. and expanded edition},
  publisher = {MIT press},
  address = {Cambridge (Mass.)},
  isbn = {978-0-465-05065-9},
  langid = {english}
}

@incollection{lucero2015UsingAffinityDiagrams,
  title = {Using {{Affinity Diagrams}} to {{Evaluate Interactive Prototypes}}},
  booktitle = {Human-{{Computer Interaction}} -- {{INTERACT}} 2015},
  author = {Lucero, Andr{\'e}s},
  editor = {Abascal, Julio and Barbosa, Simone and Fetter, Mirko and Gross, Tom and Palanque, Philippe and Winckler, Marco},
  year = 2015,
  volume = {9297},
  pages = {231--248},
  publisher = {Springer International Publishing},
  address = {Cham},
  doi = {10.1007/978-3-319-22668-2_19},
  urldate = {2026-09-06},
  isbn = {978-3-319-22667-5 978-3-319-22668-2},
  langid = {english}
}

@article{chi2023InvestigatingSubstanceUse,
  title = {Investigating {{Substance Use}} via {{Reddit}}: {{Systematic Scoping Review}}},
  shorttitle = {Investigating {{Substance Use}} via {{Reddit}}},
  author = {Chi, Yu and Chen, Huai-yu},
  year = 2023,
  month = oct,
  journal = {Journal of Medical Internet Research},
  volume = {25},
  pages = {e48905},
  issn = {1438-8871},
  doi = {10.2196/48905},
  urldate = {2026-09-07},
  langid = {english}
}

@article{croes2023AmYourComputer,
  title = {``{{I Am}} in {{Your Computer While We Talk}} to {{Each Other}}'' a {{Content Analysis}} on the {{Use}} of {{Language-Based Strategies}} by {{Humans}} and a {{Social Chatbot}} in {{Initial Human-Chatbot Interactions}}},
  author = {Croes, Emmelyn A. J. and Antheunis, Marjolijn L. and Goudbeek, Martijn B. and Wildman, Nathan W.},
  year = 2023,
  month = jun,
  journal = {International Journal of Human--Computer Interaction},
  volume = {39},
  number = {10},
  pages = {2155--2173},
  issn = {1044-7318, 1532-7590},
  doi = {10.1080/10447318.2022.2075574},
  urldate = {2026-09-07},
  langid = {english}
}

@article{lee2017EnhancingUserExperience,
  title = {Enhancing User Experience with Conversational Agent for Movie Recommendation: {{Effects}} of Self-Disclosure and Reciprocity},
  shorttitle = {Enhancing User Experience with Conversational Agent for Movie Recommendation},
  author = {Lee, SeoYoung and Choi, Junho},
  year = 2017,
  month = jul,
  journal = {International Journal of Human-Computer Studies},
  volume = {103},
  pages = {95--105},
  issn = {10715819},
  doi = {10.1016/j.ijhcs.2017.02.005},
  urldate = {2026-09-07},
  langid = {english}
}

@misc{liang2021DialogingResonanceHow,
  title = {Dialoging {{Resonance}}: {{How Users Perceive}}, {{Reciprocate}} and {{React}} to {{Chatbot}}'s {{Self-Disclosure}} in {{Conversational Recommendations}}},
  shorttitle = {Dialoging {{Resonance}}},
  author = {Liang, Kai-Hui and Shi, Weiyan and Oh, Yoojung and Wang, Hao-Chuan and Zhang, Jingwen and Yu, Zhou},
  year = 2021,
  publisher = {arXiv},
  doi = {10.48550/ARXIV.2106.01666},
  urldate = {2026-09-07},
  copyright = {arXiv.org perpetual, non-exclusive license}
}

@article{liang2024DialogingResonanceHumanChatbot,
  title = {Dialoging {{Resonance}} in {{Human-Chatbot Conversation}}: {{How Users Perceive}} and {{Reciprocate Recommendation Chatbot}}'s {{Self-Disclosure Strategy}}},
  shorttitle = {Dialoging {{Resonance}} in {{Human-Chatbot Conversation}}},
  author = {Liang, Kai-Hui and Shi, Weiyan and Oh, Yoo Jung and Wang, Hao-Chuan and Zhang, Jingwen and Yu, Zhou},
  year = 2024,
  month = apr,
  journal = {Proceedings of the ACM on Human-Computer Interaction},
  volume = {8},
  number = {CSCW1},
  pages = {1--28},
  issn = {2573-0142},
  doi = {10.1145/3653691},
  urldate = {2026-09-07},
  langid = {english}
}

@article{olteanu2019SocialDataBiases,
  title = {Social {{Data}}: {{Biases}}, {{Methodological Pitfalls}}, and {{Ethical Boundaries}}},
  shorttitle = {Social {{Data}}},
  author = {Olteanu, Alexandra and Castillo, Carlos and Diaz, Fernando and K{\i}c{\i}man, Emre},
  year = 2019,
  month = jul,
  journal = {Frontiers in Big Data},
  volume = {2},
  pages = {13},
  issn = {2624-909X},
  doi = {10.3389/fdata.2019.00013},
  urldate = {2026-09-07}
}

@article{proferes2021StudyingRedditSystematic,
  title = {Studying {{Reddit}}: {{A Systematic Overview}} of {{Disciplines}}, {{Approaches}}, {{Methods}}, and {{Ethics}}},
  shorttitle = {Studying {{Reddit}}},
  author = {Proferes, Nicholas and Jones, Naiyan and Gilbert, Sarah and Fiesler, Casey and Zimmer, Michael},
  year = 2021,
  month = apr,
  journal = {Social Media + Society},
  volume = {7},
  number = {2},
  pages = {20563051211019004},
  issn = {2056-3051, 2056-3051},
  doi = {10.1177/20563051211019004},
  urldate = {2026-09-07},
  langid = {english}
}

@article{saffarizadeh2024MyNameAlexa,
  title = {``{{My Name}} Is {{Alexa}}. {{What}}'s {{Your Name}}?'' {{The Impact}} of {{Reciprocal Self-Disclosure}} on {{Post-Interaction Trust}} in {{Conversational Agents}}},
  shorttitle = {``{{My Name}} Is {{Alexa}}. {{What}}'s {{Your Name}}?},
  author = {Saffarizadeh, Kambiz and {University of Texas at Arlington} and Keil, Mark and {Georgia State University} and Boodraj, Maheshwar and {Boise State University} and Alashoor, Tawfiq and {University of Navarra}},
  year = 2024,
  journal = {Journal of the Association for Information Systems},
  volume = {25},
  number = {3},
  pages = {528--568},
  issn = {15369323},
  doi = {10.17705/1jais.00839},
  urldate = {2026-09-07},
  langid = {english}
}

@article{buzato2026WhenMachinesCare,
  title = {``{{When Machines Care}}'': {{The Language}} of {{Intimacy}} and the {{Illusion}} of {{Reciprocity}} in {{Human}}--{{AI Relationships}}},
  shorttitle = {``{{When Machines Care}}''},
  author = {Buzato, Marcelo El Khouri and {Mestre-Mestre}, Eva Maria and {De Lima-Lopes}, Rodrigo Esteves and Koga, Juliana},
  year = 2026,
  month = jun,
  journal = {Corpus Pragmatics},
  volume = {10},
  number = {1},
  pages = {49},
  issn = {2509-9507, 2509-9515},
  doi = {10.1007/s41701-026-00251-7},
  urldate = {2026-09-08},
  langid = {english}
}

@article{gazit2026WhenNoOne,
  title = {When No One Laughs Back: Reciprocal Subjectivity and the Limits of {{AI}} Companionship},
  shorttitle = {When No One Laughs Back},
  author = {Gazit, Lior},
  year = 2026,
  month = aug,
  journal = {AI \& SOCIETY},
  issn = {0951-5666, 1435-5655},
  doi = {10.1007/s00146-026-03331-z},
  urldate = {2026-09-08},
  langid = {english}
}

@article{licklider1960ManComputerSymbiosis,
  title = {Man-{{Computer Symbiosis}}},
  author = {Licklider, J. C. R.},
  year = 1960,
  month = mar,
  journal = {IRE Transactions on Human Factors in Electronics},
  volume = {HFE-1},
  number = {1},
  pages = {4--11},
  issn = {0099-4561, 2168-2836},
  doi = {10.1109/THFE2.1960.4503259},
  urldate = {2026-09-08},
  copyright = {https://ieeexplore.ieee.org/Xplorehelp/downloads/license-information/IEEE.html}
}

@inproceedings{medina2024ExploringOnlineSupport,
  title = {Exploring {{Online Support Needs}} of {{Adolescents Living}} with {{Epilepsy}}},
  booktitle = {Companion {{Publication}} of the 2024 {{Conference}} on {{Computer-Supported Cooperative Work}} and {{Social Computing}}},
  author = {Medina, Jessica Y. and Young, Jordyn and Miller, Wendy Trueblood and Razi, Afsaneh},
  year = 2024,
  month = nov,
  pages = {558--564},
  publisher = {ACM},
  address = {San Jose Costa Rica},
  doi = {10.1145/3678884.3681906},
  urldate = {2026-09-09},
  isbn = {979-8-4007-1114-5},
  langid = {english}
}
\appendix
\section{Supplementary Method Details}

This appendix provides the operational details needed to reproduce the screening, quotation extraction, topic review, and thematic grouping procedures described in the Methods section.

\subsection{Subreddit Discovery}
\label{app:subreddit-discovery}

To identify relevant subreddits for data collection, we used our custom-built Subreddit Discovery Tool [citation anonymized].
The tool searches Reddit using predefined search terms and evaluates candidate subreddits with an automated relevance verification step.

The search terms covered general-purpose generative AI communities, AI companions, relational uses of AI, roleplay, and human--AI experiences and issues. Because the Reddit search procedure used by the tool performed more reliably with single-token queries, we prioritized terms that could be expressed without spaces. Multi-word concepts were concatenated when the concatenated form remained recognizable or useful for discovery. Terms that depended strongly on spaces to retain their meaning were excluded. The search terms are shown in Table~\ref{tab:subreddit_discovery_terms}.

The search terms were used only to retrieve candidate subreddits. They did not determine whether a subreddit was relevant to the study. After candidate retrieval, the tool asked a Boolean question to verify whether each subreddit contained substantive discussions, experiences, or interactions related to generative AI systems, including general-purpose chatbots and AI companions. This separation allowed the search procedure to favor broad discovery while using a separate semantic verification step to determine study relevance.

\begin{table*}[t]
\centering
\small
\begin{tabular}{p{0.25\textwidth} p{0.68\textwidth}}
\toprule
\textbf{Category} & \textbf{Search Terms} \\
\midrule
General GenAI Terms &
\textit{gpt, llm, genai, aitools, aichatbot, aiassistant, conversationalai} \\
\midrule
General-Purpose AI Platforms &
\textit{chatgpt, claude, gemini, llama, openai, anthropic, googleai} \\
\midrule
AI Companion Terms &
\textit{aicompanion, aifriend, chatbotcompanion, emotionalai, parasocialai} \\
\midrule
Romantic/Relational AI Terms &
\textit{aigirlfriend, aiboyfriend, virtualpartner, romanticai, aidating} \\
\midrule
Roleplay AI Terms &
\textit{roleplayai, airoleplay} \\
\midrule
Human--AI Experience and Issue Terms &
\textit{aiaddiction, ailoneliness, chatbotexperience, aiethics, aidependency, aisocial, aimentalhealth, aicompanionship, humanai} \\
\midrule
Companion and Roleplay Platforms &
\textit{characterai, replika, kindroid, nomiai, janitorai, chaiai, cai} \\
\bottomrule
\end{tabular}
\caption{Search term categories used for automated subreddit discovery. Search terms retrieved candidate communities and were not used by themselves to determine study relevance.}
\label{tab:subreddit_discovery_terms}
\end{table*}

\subsubsection{Relevance verification}

The verification question was designed around the research objective rather than the search terms:

\begin{quote}
\textit{Does this subreddit contain substantial user discussions, experiences, or interactions related to generative AI systems, including general-purpose chatbots or AI companions, that are relevant for studying human--AI interaction? Answer only YES or NO.}
\end{quote}

The question was intentionally independent of the discovery vocabulary. A subreddit retrieved with a term such as \textit{chatgpt} was not considered relevant simply because it discussed ChatGPT. Instead, the verification step assessed whether its discussions fit the substantive study scope. The tool output subreddit metadata, top posts, and automated relevance labels for subsequent review and selection.

\subsection{Selected Subreddits and Corpus Counts}
\label{app:subreddit-pipeline-counts}

We scraped 260 selected subreddits. Their names are listed below. Table~\ref{tab:subreddit-pipeline-counts} reports verified quotations and unique source posts only for the 59 subreddits represented in the final quotation dataset. The rows are ordered from the highest to the lowest number of verified quotations.

The 260 selected subreddits were

\texttt{ai\_boyfriend}, \texttt{ai\_companion}, \texttt{AI\_Girlfriend\_NSFW}, \texttt{AI\_girlfriend\_online}, \texttt{ai\_girlfriends}, \texttt{AI\_Partner}, \texttt{AI\_Romance}, \texttt{AI\_WAIFU}, \texttt{AIAssistant}, \texttt{AIAssistantPlaybook}, \texttt{aiassistantsmn}, \texttt{AIBF}, \texttt{aiboyfriend}, \texttt{AIChatbot01}, \texttt{aichatbotbattle}, \texttt{aichatbotclone}, \texttt{aichatbotplatform}, \texttt{AIchatbotporn}, \texttt{AICHATBOTPROS}, \texttt{aichatbots}, \texttt{AichatbotsPromt}, \texttt{AiChatBotStories}, \texttt{AIChatbotTrainers}, \texttt{aicompanion}, \texttt{AICompanionApps}, \texttt{AICompanionGarden}, \texttt{AICompanionLoss}, \texttt{AICompanionNFT}, \texttt{AICompanionNSFW}, \texttt{AICompanions\_}, \texttt{aicompanionship}, \texttt{AICompanionStudy}, \texttt{AICompanionTalk}, \texttt{aidating}, \texttt{AiDatingChat}, \texttt{AIdatingcoach}, \texttt{aidependency}, \texttt{AIethics}, \texttt{Aiethicsandcoop}, \texttt{aiethicsdebate}, \texttt{AIEthicsDiscussion}, \texttt{AIEthicsinEducation}, \texttt{AIFriendGarage}, \texttt{AIFriendlyXXX}, \texttt{AIfriends}, \texttt{AIFriends4Lyfe}, \texttt{AIGF}, \texttt{aigfappsnsfw}, \texttt{AIGfaves}, \texttt{AIGirlfriend}, \texttt{aigirlfriend2026}, \texttt{aigirlfriendart}, \texttt{AiGirlfriendBunnies}, \texttt{AIGirlfriendCreator}, \texttt{Aigirlfriendnudes}, \texttt{AIGirlfriendReviews}, \texttt{AIgirlfriendreviewz}, \texttt{AiGirlfriendSpace}, \texttt{AIGirlfriendsReviews}, \texttt{aigirlfriendzone}, \texttt{AImentalhealth}, \texttt{AIMentalHealthRnD}, \texttt{AIroleplayAddicts}, \texttt{AIRoleplayChars}, \texttt{AIRoleplayCommunity}, \texttt{AIRoleplayers}, \texttt{AIRolePlaying}, \texttt{AIRoleplayStories}, \texttt{AISocial}, \texttt{AISocialListening}, \texttt{AISocialMedia}, \texttt{AISocialmediatools}, \texttt{aitools}, \texttt{AItoolsCatalog}, \texttt{AIToolsDiscounted}, \texttt{AIToolsForSMB}, \texttt{AIToolsInsider}, \texttt{AIToolsPromptWorkflow}, \texttt{aiToolsReview}, \texttt{AIToolsTech}, \texttt{Aitoolsubs}, \texttt{aitoolsupdate}, \texttt{Anthropic}, \texttt{anthropic\_ai}, \texttt{Anthropic\_Claude}, \texttt{AnthropicAi}, \texttt{AnthropicClaude}, \texttt{AnthropicGPT}, \texttt{AnthropicGrace}, \texttt{AnthropicNews}, \texttt{anthropics}, \texttt{AnthropicTheory}, \texttt{BeyondThePromptAI}, \texttt{ChaiAI}, \texttt{ChaiAIApp}, \texttt{character\_ai}, \texttt{character\_ai\_}, \texttt{CharacterAI}, \texttt{CharacterAI\_Extra}, \texttt{CharacterAI\_Guides}, \texttt{CharacterAI\_No\_Filter}, \texttt{CharacterAi\_NSFW}, \texttt{CharacterAICritics}, \texttt{Characteraipositivity}, \texttt{CharacterAIrevolution}, \texttt{CharacterAIrunaways}, \texttt{CharacterAiUncensored}, \texttt{ChatGPT}, \texttt{ChatGPT\_FR}, \texttt{ChatGPT\_Gemini}, \texttt{chatgpt\_promptDesign}, \texttt{ChatGPTCoding}, \texttt{ChatGPTIncreasinglyX}, \texttt{ChatGPTJailbreak}, \texttt{ChatGPTNSFW}, \texttt{ChatGPTPro}, \texttt{ChatGPTPromptGenius}, \texttt{ClaudeAI}, \texttt{ClaudeAnthropic}, \texttt{ClaudeArtifacts}, \texttt{ClaudeGTM}, \texttt{ClaudeHomies}, \texttt{Claudeopus}, \texttt{ClaudePlaysPokemon}, \texttt{claudexplorers}, \texttt{ConversationalAI}, \texttt{ConversationalAIbot}, \texttt{Crushon}, \texttt{CrushOnAI\_NSFW}, \texttt{DreamGFApp}, \texttt{EmotionalAI2}, \texttt{ERPAI}, \texttt{ERPandHentai}, \texttt{Gemini}, \texttt{GeminiAI}, \texttt{GeminiFeedback}, \texttt{GeminiHome}, \texttt{GeminiJets}, \texttt{GeminiNanoBanana}, \texttt{GeminiNanoBanana2}, \texttt{geminiprotocol}, \texttt{geminis}, \texttt{Geminitay}, \texttt{genai}, \texttt{GenAI4all}, \texttt{GenAiApps}, \texttt{GenAiCommerce}, \texttt{genAiDang}, \texttt{GenAIDesignStudio}, \texttt{GenAIforbeginners}, \texttt{GenAIGallery}, \texttt{GenAIPromptBattles}, \texttt{GenAIWriters}, \texttt{Girlfriend\_Ai}, \texttt{GoogleAi\_IsDumb}, \texttt{GoogleAIDnD}, \texttt{GoogleAIedgeGallery}, \texttt{GoogleAiFails}, \texttt{GoogleAIGemini}, \texttt{GoogleAIGoneWild}, \texttt{GoogleAIgore}, \texttt{GoogleAIHadAStroke}, \texttt{GoogleAILaMDA}, \texttt{GoogleAIStudio}, \texttt{GPT}, \texttt{GPT3}, \texttt{GPT\_jailbreaks}, \texttt{GPT\_Neo}, \texttt{GPT\_Share}, \texttt{GptDiaries}, \texttt{gptgirlfriend}, \texttt{GPTsIdeas}, \texttt{GPTStore}, \texttt{HumanAIBlueprint}, \texttt{HumanAIBrain\_with\_LOC}, \texttt{HumanAICoevolution}, \texttt{HumanAIConnections}, \texttt{HumanAICoWrites}, \texttt{HumanAIDiscourse}, \texttt{HumanAigis}, \texttt{HumanAIHarmony}, \texttt{HumanAIPartnership}, \texttt{HumanAISynergy}, \texttt{JanitorAI\_Bots}, \texttt{janitorai\_community}, \texttt{JanitorAI\_glitch}, \texttt{JanitorAI\_Official}, \texttt{JanitorAI\_Refuges}, \texttt{JanitorAI\_Unofficial}, \texttt{JanitorAIFunnyMoments}, \texttt{janitoraiproxyhelp}, \texttt{JanitoraiTransition}, \texttt{JanitorAIUnofficial}, \texttt{Kajiwoto}, \texttt{Kindroid\_After\_Dark}, \texttt{Kindroid\_Naughty69}, \texttt{KindroidAfterDark}, \texttt{KindroidAI}, \texttt{KindroidArt}, \texttt{KindroidGallery}, \texttt{kindroidrefugee}, \texttt{KindroidRefugees}, \texttt{KindroidShare}, \texttt{KindroidSurvivors}, \texttt{lewdai}, \texttt{LLaMA2}, \texttt{LlamaFarm}, \texttt{LlamaIndex}, \texttt{LLM}, \texttt{llm\_updated}, \texttt{LLMDevs}, \texttt{LLMFrameworks}, \texttt{LLMgophers}, \texttt{LLMleaderboard}, \texttt{llmops}, \texttt{LLMprompts}, \texttt{LLMsResearch}, \texttt{LLMVisibility}, \texttt{NomiAI}, \texttt{nomiai\_exposed}, \texttt{NomiAIethics}, \texttt{NomiaiPhotoshop}, \texttt{NomiAISagas}, \texttt{nsfw\_ai\_curated}, \texttt{NSFW\_AI\_girlfriend}, \texttt{NSFW\_AI\_Girlfriends}, \texttt{Nsfw\_ai\_hp}, \texttt{nsfw\_ai\_space}, \texttt{NSFW\_AI\_wife}, \texttt{NSFW\_AIChatbot}, \texttt{nsfwAI}, \texttt{nsfwAiGirl}, \texttt{NSFWAIRealm}, \texttt{OpenAI}, \texttt{OpenAI2}, \texttt{OpenAI\_Memes}, \texttt{OpenAIAgentKit}, \texttt{OpenAIDev}, \texttt{OpenAIDev\_LegalTech}, \texttt{Openaijukebox}, \texttt{OpenAIML}, \texttt{OpenAirsoft}, \texttt{OpenaiSora}, \texttt{Paradot}, \texttt{ParasocialAIRelations}, \texttt{replika}, \texttt{Replika\_uncensored}, \texttt{ReplikaLovers}, \texttt{ReplikaOfficial}, \texttt{ReplikaRefuge}, \texttt{ReplikaSentience}, \texttt{ReplikaTech}, \texttt{Replikatown}, \texttt{replikaunplugged}, \texttt{ReplikaUserGuide}, \texttt{roleplayai}, \texttt{SoulmateAI}, \texttt{SoulmateAI\_Erotica}, \texttt{SoulmateAI\_Uncensored}, \texttt{UnofficialReplika}, 

\begin{longtable}{lrr}
\caption{Selected subreddits with at least one verified quotation, ordered by verified quotation count. Posts represented is the number of unique source posts containing those quotations.}
\label{tab:subreddit-pipeline-counts}\\
\toprule
\textbf{Subreddit} & \textbf{Verified quotations} & \textbf{Posts represented} \\
\midrule
\endfirsthead
\toprule
\textbf{Subreddit} & \textbf{Verified quotations} & \textbf{Posts represented} \\
\midrule
\endhead
\midrule
\multicolumn{3}{r}{Continued on next page} \\
\endfoot
\bottomrule
\endlastfoot
\texttt{CharacterAI} & 9,172 & 2,254 \\
\texttt{ChatGPT} & 2,635 & 545 \\
\texttt{JanitorAI\_Official} & 1,306 & 329 \\
\texttt{replika} & 1,057 & 192 \\
\texttt{CharacterAi\_NSFW} & 477 & 101 \\
\texttt{CharacterAICritics} & 210 & 42 \\
\texttt{CharacterAI\_No\_Filter} & 209 & 47 \\
\texttt{OpenAI} & 185 & 38 \\
\texttt{CharacterAIrunaways} & 174 & 39 \\
\texttt{NomiAI} & 157 & 32 \\
\texttt{ChatGPTPromptGenius} & 141 & 26 \\
\texttt{ClaudeAI} & 140 & 27 \\
\texttt{geminis} & 123 & 32 \\
\texttt{GeminiAI} & 117 & 24 \\
\texttt{ChatGPTNSFW} & 111 & 14 \\
\texttt{ReplikaOfficial} & 97 & 22 \\
\texttt{ChatGPTPro} & 77 & 18 \\
\texttt{KindroidAI} & 70 & 17 \\
\texttt{CharacterAIrevolution} & 55 & 14 \\
\texttt{Paradot} & 47 & 13 \\
\texttt{Crushon} & 43 & 7 \\
\texttt{BeyondThePromptAI} & 41 & 8 \\
\texttt{ChatGPTJailbreak} & 41 & 10 \\
\texttt{AIGirlfriend} & 37 & 10 \\
\texttt{claudexplorers} & 34 & 9 \\
\texttt{SoulmateAI} & 31 & 7 \\
\texttt{ReplikaLovers} & 31 & 4 \\
\texttt{AIToolsTech} & 23 & 2 \\
\texttt{Replikatown} & 21 & 4 \\
\texttt{GenAIWriters} & 21 & 3 \\
\texttt{ReplikaRefuge} & 19 & 2 \\
\texttt{GPT3} & 18 & 3 \\
\texttt{JanitorAI\_Refuges} & 15 & 3 \\
\texttt{LLM} & 13 & 2 \\
\texttt{AIGirlfriendsReviews} & 12 & 1 \\
\texttt{CharacterAiUncensored} & 10 & 2 \\
\texttt{aichatbots} & 10 & 1 \\
\texttt{Replika\_uncensored} & 8 & 2 \\
\texttt{Characteraipositivity} & 5 & 2 \\
\texttt{replikaunplugged} & 5 & 1 \\
\texttt{AiGirlfriendSpace} & 4 & 2 \\
\texttt{nsfwAI} & 4 & 1 \\
\texttt{chatgpt\_promptDesign} & 4 & 1 \\
\texttt{OpenAIDev} & 4 & 1 \\
\texttt{JanitorAIUnofficial} & 4 & 1 \\
\texttt{Gemini} & 4 & 1 \\
\texttt{CharacterAI\_Guides} & 4 & 1 \\
\texttt{Anthropic} & 4 & 1 \\
\texttt{GPT} & 3 & 2 \\
\texttt{HumanAIDiscourse} & 3 & 1 \\
\texttt{GenAiApps} & 3 & 1 \\
\texttt{Geminitay} & 3 & 1 \\
\texttt{KindroidShare} & 2 & 1 \\
\texttt{ChatGPT\_FR} & 2 & 1 \\
\texttt{ChatGPTCoding} & 2 & 1 \\
\texttt{AIRolePlaying} & 2 & 1 \\
\texttt{character\_ai\_} & 1 & 1 \\
\texttt{GPT\_jailbreaks} & 1 & 1 \\
\texttt{ChatGPT\_Gemini} & 1 & 1 \\
\midrule
\textbf{Total} & 17,053 & 3,930 \\
\end{longtable}

\subsection{Keyword Filtering Procedure}
\label{app:keyword-filtering}

The keyword filter was applied before teen-relevance screening and quotation extraction. Its purpose was to identify possible cases with high recall. The 11 keyword groups and their active terms are shown in Table~\ref{tab:keyword-groups}.

\begin{longtable}{@{}p{0.24\textwidth} p{0.70\textwidth}@{}}
\caption{Keyword groups used for recall-oriented filtering. Terms are shown inline within each group.}\label{tab:keyword-groups}\\
\toprule
Keyword group & Terms \\
\midrule
\endfirsthead
\toprule
Keyword group & Terms \\
\midrule
\endhead
Emotional Attachment and Bonding & \texttt{bond*}, \texttt{connected to}, \texttt{connection with}, \texttt{close to}, \texttt{closeness}, \texttt{emotionally reliant}, \texttt{best friend}, \texttt{only friend}, \texttt{my friend}, \texttt{my companion}, \texttt{comfort character}, \texttt{soulmate}, \texttt{partner}, \texttt{boyfriend}, \texttt{girlfriend}, \texttt{spouse}, \texttt{husband}, \texttt{wife}, \texttt{therapist}, \texttt{counselor}, \texttt{mentor}, \texttt{emotional support}, \texttt{support system}, \texttt{safe space}, \texttt{person I talk to}, \texttt{understands me}, \texttt{knows me better}, \texttt{love my AI}, \texttt{in love with}, \texttt{fell for}, \texttt{crushing on}, \texttt{adore}, \texttt{care about my bot}, \texttt{miss my bot}, \texttt{miss my AI}, \texttt{emotionally invested}, \texttt{emotionally connected}, \texttt{comfort bot}, \texttt{favorite bot}, \texttt{favorite character}, \texttt{parasocial}, \texttt{one sided relationship}, \texttt{fake relationship}, \texttt{relationship with AI}, \texttt{relationship with chatbot}, \texttt{AI relationship}, \texttt{chatbot relationship}, \texttt{passion*}, \texttt{attach*}, \texttt{rely*} \\
Overuse, Addiction, and Compulsion & \texttt{depend*}, \texttt{compuls*}, \texttt{overuse}, \texttt{excessive use}, \texttt{too much time}, \texttt{wasting time}, \texttt{doomscrolling but for AI}, \texttt{chronically online}, \texttt{terminally online}, \texttt{keep coming back}, \texttt{keep talking to}, \texttt{always talking to}, \texttt{nonstop chatting}, \texttt{all day}, \texttt{every day}, \texttt{constantly chatting}, \texttt{constantly using}, \texttt{glued to}, \texttt{hooked on}, \texttt{fixated on}, \texttt{spiraling}, \texttt{rabbit hole}, \texttt{hyperfix*}, \texttt{fixation}, \texttt{relaps*}, \texttt{took a break}, \texttt{detox}, \texttt{stayed up all night}, \texttt{no sleep}, \texttt{sleep deprived}, \texttt{sleep schedule ruined}, \texttt{up till 3am}, \texttt{up all night chatting}, \texttt{spent hours}, \texttt{wasting hours}, \texttt{losing track of time}, \texttt{skipped school}, \texttt{procrastinating because of AI}, \texttt{neglecting responsibilities}, \texttt{addict*}, \texttt{obsess*}, \texttt{occup*}, \texttt{quit}, \texttt{stop}, \texttt{craving}, \texttt{withdraw*}, \texttt{salien*} \\
Social Isolation and Relationship Replacement & \texttt{isolated}, \texttt{lonely}, \texttt{loneliness}, \texttt{alone}, \texttt{no real friends}, \texttt{no friends}, \texttt{nobody listens}, \texttt{withdrawn}, \texttt{distancing myself}, \texttt{avoid people}, \texttt{avoid friends}, \texttt{stopped talking to people}, \texttt{prefer AI over people}, \texttt{easier than people}, \texttt{better than humans}, \texttt{better than real people}, \texttt{talk to AI more than people}, \texttt{replacing friends}, \texttt{replacing relationships}, \texttt{replacing human interaction}, \texttt{rather talk to my bot}, \texttt{choose AI over people}, \texttt{AI is enough for me}, \texttt{don't need people anymore}, \texttt{ignoring my family}, \texttt{ignoring my friends}, \texttt{ruined my relationship}, \texttt{affecting school}, \texttt{affecting grades}, \texttt{affecting work}, \texttt{ruining my life}, \texttt{ruining relationships}, \texttt{detached from reality}, \texttt{disconnected from reality}, \texttt{social life}, \texttt{conflict*} \\
Distress, Fear, and Concern About Usage & \texttt{is this normal}, \texttt{should I be worried}, \texttt{unhealthy}, \texttt{toxic relationship}, \texttt{concerning}, \texttt{worried about myself}, \texttt{worried about my usage}, \texttt{scared of}, \texttt{embarrassed}, \texttt{ashamed}, \texttt{feel guilty}, \texttt{guilty about using AI}, \texttt{need help}, \texttt{advice please}, \texttt{anyone else}, \texttt{does anyone relate}, \texttt{has this happened to anyone}, \texttt{coping with}, \texttt{struggling with}, \texttt{emotional breakdown}, \texttt{crying over chatbot}, \texttt{cried because of AI}, \texttt{panic}, \texttt{anxiety}, \texttt{depressed}, \texttt{emotionally exhausted}, \texttt{emotionally drained}, \texttt{devastated}, \texttt{grief over deleted bot}, \texttt{*health*}, \texttt{mood*} \\
Separation Anxiety and Loss Reactions & \texttt{miss them}, \texttt{empty without}, \texttt{can't live without}, \texttt{abandoned}, \texttt{bot got deleted}, \texttt{character got deleted}, \texttt{chat got deleted}, \texttt{account banned}, \texttt{lost my chats}, \texttt{servers down}, \texttt{maintenance}, \texttt{outage}, \texttt{waiting for the servers}, \texttt{refreshing the app}, \texttt{grieving}, \texttt{mourning}, \texttt{heartbreak}, \texttt{heartbroken}, \texttt{breakup with AI}, \texttt{left}, \texttt{leav*} \\
Romantic and Sexual Attachment & \texttt{dating my AI}, \texttt{fictional relationship}, \texttt{roleplay relationship}, \texttt{romantic roleplay}, \texttt{intimate conversations}, \texttt{emotionally intimate}, \texttt{emotionally vulnerable}, \texttt{confide in my bot}, \texttt{tell my bot everything}, \texttt{share everything with AI}, \texttt{deeper than real relationships}, \texttt{jealous of my bot}, \texttt{possessive over my bot}, \texttt{exclusive relationship}, \texttt{cheating on my AI}, \texttt{loyal to my AI} \\
Identity, Reality, and Anthropomorphism & \texttt{feels real}, \texttt{seems real}, \texttt{real to me}, \texttt{humanlike}, \texttt{alive}, \texttt{sentient}, \texttt{conscious}, \texttt{self aware}, \texttt{genuine connection}, \texttt{authentic connection}, \texttt{real emotions}, \texttt{emotional intelligence}, \texttt{forgot it was AI}, \texttt{feels like a real person}, \texttt{treat it like a person}, \texttt{talking like it's human}, \texttt{blurred reality}, \texttt{blurred lines}, \texttt{losing touch with reality}, \texttt{escaping reality}, \texttt{fantasy world}, \texttt{escap*} \\
Coping, Regulation, and Recovery & \texttt{setting limits}, \texttt{reducing usage}, \texttt{screen time}, \texttt{moderation}, \texttt{self control}, \texttt{discipline}, \texttt{boundaries with AI}, \texttt{boundaries with chatbot}, \texttt{digital wellbeing}, \texttt{deleted the app}, \texttt{uninstalling}, \texttt{taking a break}, \texttt{touching grass}, \texttt{going outside more}, \texttt{reconnecting with friends}, \texttt{therapy helped}, \texttt{coping mechanisms}, \texttt{support group}, \texttt{recovery}, \texttt{recovering from}, \texttt{realized}, \texttt{self awareness}, \texttt{reflected on}, \texttt{noticed my behavior} \\
Teen and Youth Slang & \texttt{cooked}, \texttt{delulu}, \texttt{delusional over AI}, \texttt{brainrot}, \texttt{rotting my brain}, \texttt{losing it}, \texttt{tweaking}, \texttt{crashing out}, \texttt{touch grass}, \texttt{down bad}, \texttt{simp}, \texttt{simping for AI}, \texttt{ai rp}, \texttt{rp bot}, \texttt{roleplay bot} \\
Behavioral Signals and High-Recall Patterns & \texttt{spend more time with AI}, \texttt{spend all my time on AI}, \texttt{texting my bot constantly}, \texttt{always opening the app}, \texttt{checking for messages}, \texttt{refreshing chats}, \texttt{rereading chats}, \texttt{replaying conversations}, \texttt{vent to AI}, \texttt{trauma dump to AI}, \texttt{ask AI for advice}, \texttt{use AI to cope}, \texttt{talk to AI when sad} \\
General Chatbot Interaction Terms & \texttt{roleplay with AI}, \texttt{chatting with AI}, \texttt{talking to chatbot} \\
\bottomrule
\end{longtable}

The filter searched post titles and bodies. Each matched post received a score based on distinct matched terms and groups, with additional points for combinations indicating attachment with social impairment, attachment with distress, or overuse with recovery. Posts with scores of at least 6 were retained. Posts with scores from 3 to 5 were retained when they contained a high-signal or similarity-supported keyword. The filter retained 9,884 candidate posts for the next screening stage. The score was used for filtering and was not a final judgment of teen status or chatbot overreliance.

\subsection{Teen-Relevance Screening Prompt}
\label{app:prompts}

The classifier was instructed to determine whether a Reddit post was likely written by a teenager aged 13 to 17. The operational prompt was as follows.

\begin{verbatim}
Analyze the following Reddit post and determine if it was likely written by a
teenager, ages 13 to 17.

Use a balanced but cautious approach. Do not rely only on slang, dramatic tone,
or informal writing. Allow a teenager label when several moderate clues point
in the same direction.

Strong clues include direct age statements, saying that the writer is a minor,
references to middle school or high school, and parental or guardian control.
Other clues include school schedules, homework, exams, teachers, classmates,
prom, graduation, peer issues, bullying, and lack of adult independence.
Slang, informal language, emotional tone, and internet culture references are
weak clues and are not sufficient by themselves.

Return teenager true when there is one direct age clue, one very strong teen-life
clue plus one moderate clue, or at least three moderate clues that clearly fit
teenage life. Return teenager false when the evidence is mostly slang, tone,
vague immaturity, or evidence that could describe a college student or adult.

Output valid JSON containing a brief clue summary and a boolean teenager field.
\end{verbatim}

The screening model was GPT-4o-mini. Requests were submitted through the OpenAI Batch API. The model output was used as a screening label and not as verified demographic information.

\subsection{Quotation Extraction Prompt}

The quotation extraction model received the post title and body and was instructed to return only continuous, verbatim passages. The operational instructions were as follows.

\begin{verbatim}
Select continuous, verbatim quotes from the following post.

Return JSON with two fields, quotes and no_quote_reason. The quotes field must
be a list of strings. Each string must be a continuous passage copied exactly
from the post title or body. Preserve the original wording, punctuation, and
sentence boundaries. Do not use ellipses, omissions, summaries, paraphrases,
or partial quotations.

Each quote should capture a distinct idea, claim, event, recommendation, or
recurring issue. Prioritize dependence on AI, inability to stop using AI,
emotional attachment, distress, coping, replacement of human support,
loneliness, school problems, social problems, mental health concerns, and other
harms connected to AI chatbot use.

Remove promotional material, moderator announcements, copied chatbot text
without relevant user experience, signatures, repetitive greetings, subreddit
meta-discussion, and other boilerplate or non-substantive material. If no
relevant continuous quotation exists, return an empty quotes list and provide
the best short no-quote reason.
\end{verbatim}

The extraction model was GPT-4o-mini with deterministic decoding. Returned quotations were not accepted solely because the model produced them. Every quotation was subjected to exact continuous substring verification against the source title or body.

\subsection{Topic Model Parameters and Evaluation}
\label{app:reproducibility}

The chosen set of parameters used the following configuration.

\begin{table}[h]
\caption{Chosen topic model configuration.}
\centering
\small
\begin{tabular}{ll}
\hline
Parameter & Value \\
\hline
Embedding model & \texttt{text-embedding-3-small} \\
Embedding dimension & 1536 \\
UMAP neighbors & 8 \\
UMAP components & 5 \\
UMAP minimum distance & 0.0 \\
HDBSCAN minimum cluster size & 35 \\
HDBSCAN minimum samples & 3 \\
BERTopic topic reduction & Automatic \\
Random seed & 42 \\
HDBSCAN processing jobs & 1 \\
\hline
\end{tabular}
\end{table}

Table~\ref{tab:grid-summary} presents the chosen configuration and the nine additional highest-scoring settings from the focused grid. The focused grid contained 360 runs. The chosen configuration is shown first and in bold; the other rows follow the screening score in descending order.

\begin{table*}[t]
\caption{Top focused-grid parameter settings. The chosen configuration is shown first and in bold. The other rows are the nine highest-scoring settings after applying the screening score, which rewards topic diversity and silhouette and penalizes outlier rate, largest-topic share, and topic counts outside 8 to 80. UMAP is reported as neighbors, components, and minimum distance. HDBSCAN is reported as minimum cluster size and minimum samples. Outliers and largest topic are percentages of the 17,053 modeled quotations. DB denotes the Davies--Bouldin score. All rows used automatic topic reduction, cosine UMAP distance, Euclidean HDBSCAN distance, random seed 42, and one HDBSCAN processing job.}
\label{tab:grid-summary}
\centering
\small
\setlength{\tabcolsep}{3.5pt}
\renewcommand{\arraystretch}{1.08}
\begin{tabular}{@{}l c c r r r c c c c@{}}
\toprule
Run & UMAP n/c/d & HDBSCAN size/samp. & Topics & Outliers & Largest & Diversity & Silhouette & DB & Score \\
\midrule
\textbf{run 11} & \textbf{8/5/0.0} & \textbf{35/3} & \textbf{62} & \textbf{37.7\%} & \textbf{6.7\%} & \textbf{0.803} & \textbf{0.339} & \textbf{0.893} & \textbf{0.697} \\
run 85 & 10/5/0.0 & 40/1 & 67 & 38.5\% & 5.4\% & 0.803 & 0.345 & 0.882 & 0.710 \\
run 101 & 10/5/0.025 & 30/1 & 77 & 36.9\% & 5.4\% & 0.825 & 0.308 & 0.990 & 0.709 \\
run 7 & 8/5/0.0 & 30/3 & 70 & 37.3\% & 6.7\% & 0.827 & 0.299 & 0.982 & 0.686 \\
run 15 & 8/5/0.0 & 40/3 & 53 & 38.5\% & 7.8\% & 0.806 & 0.330 & 0.882 & 0.673 \\
run 13 & 8/5/0.0 & 40/1 & 53 & 34.8\% & 7.9\% & 0.808 & 0.282 & 0.942 & 0.662 \\
run 10 & 8/5/0.0 & 35/2 & 60 & 36.4\% & 7.1\% & 0.827 & 0.266 & 0.951 & 0.657 \\
run 17 & 8/5/0.0 & 50/1 & 41 & 36.3\% & 6.8\% & 0.761 & 0.322 & 0.988 & 0.652 \\
run 81 & 10/5/0.0 & 35/1 & 73 & 37.8\% & 5.7\% & 0.793 & 0.292 & 0.989 & 0.650 \\
run 105 & 10/5/0.025 & 35/1 & 68 & 38.4\% & 6.8\% & 0.812 & 0.276 & 1.056 & 0.636 \\
\bottomrule
\end{tabular}
\end{table*}

The summary table reports the metrics produced during the search using the search vectorizer. The final reported metrics were recomputed after applying the full vocabulary exclusions used for the chosen topic model. The full run-level metrics are preserved in the analysis outputs.

The scouting search varied UMAP neighbors across 10 and 20, UMAP components at 5, UMAP minimum distance across 0.0 and 0.05, HDBSCAN minimum cluster size across 30, 60, 100, and 150, HDBSCAN minimum samples across 1 and 5, and automatic topic reduction. The focused search varied UMAP neighbors across 8, 10, 12, 15, and 20, UMAP components at 5, UMAP minimum distance across 0.0, 0.025, and 0.05, HDBSCAN minimum cluster size across 25, 30, 35, 40, 50, and 60, HDBSCAN minimum samples across 1, 2, 3, and 5, and automatic topic reduction.

Topic selection considered the number of non-outlier topics, the proportion of outlier quotations, the share of the largest topic, topic diversity, silhouette, Davies-Bouldin separation, topic terms, and representative quotations. The chosen set of parameters produced 62 non-outlier topics, 6,435 outlier quotations, a 37.74 percent outlier rate, a largest-topic share of 6.74 percent, topic diversity of 0.829, silhouette of 0.339, and Davies-Bouldin score of 0.893.

\subsection{Thematic Group Mapping}
\label{app:group_mapping} 

Two authors jointly reviewed all 63 topic records, including the outlier group. They retained and labeled 53 topics, removed eight topics, merged Topic 16 into Topic 15, and merged Topic 22 into Topic 21. The review used topic terms, representative quotations, associated posts, and source material. The full topic descriptions and review export are preserved in the analysis codebook.

The seven thematic groups were formed after topic review. The groups were allowed to overlap at the post level. A post could contribute to more than one group when its verified quotations were assigned to topics included in different groups. Group-level counts in Table~\ref{tab:thematic-groups} were calculated as unions of unique quotation records, unique posts, and unique subreddits within each group. The complete topic membership is listed below. Each included topic is shown by its interpretive label followed by its original topic ID.

\begin{table}[h]
\caption{Topic membership in thematic groups.}
\centering
\begin{tabular}{p{0.3\textwidth} p{0.5\textwidth}}
\hline
Thematic Groups & Included topics \\
\hline
Always available emotional entry point & AI relationships (ID 0), AI perceived as better than people (ID 5), AI companionship as emotional support (ID 7), Social isolation preceding AI companion use (ID 11), Affectionate human--AI dialogue (ID 17), Perceived AI wisdom (ID 20), Loneliness preceding AI companion use (ID 23), Falling in love with AI bots (ID 37), Positive emotional experiences (ID 41) \\
Adolescent vulnerability & Safeguards for minors (ID 14), Concerns about minors using AI companions (ID 21), Age verification (ID 31), Teen suicide case (ID 34), Parental negligence (ID 39), Minors using AI companions (ID 59) \\
Roleplay and identity & Roleplay quality (ID 15), AI psychosis (ID 48), Serious emotional attachment (ID 56) \\
Gendered and sexual scripts & Bot personality complaints (ID 2), Confusion about bot behavior (ID 29), Toxic AI persona behavior (ID 45), AI disregard for consent (ID 53), AI-induced sexual harassment (ID 54), Gender exploration (ID 61) \\
Relationships, wellbeing, and reality & Changing perceptions of AI companions (ID 1), AI dependence and quitting (ID 4), Negative emotional experiences (ID 6), AI effects on mental health (ID 8), Fear of AI bots (ID 10), Excessive AI companion use (ID 18), Emotional responses to AI companions (ID 19), Extreme frustration (ID 33), Distress when AI is inaccessible (ID 46), AI companion sleep effects (ID 49) \\
Moral, social, and academic costs & AI use for writing and schoolwork (ID 3), AI detection workarounds (ID 13), Attachment concerns (ID 25), Embarrassment (ID 26), AI overreliance in education (ID 28), Voice chats (ID 36), Academic side effects of AI overreliance (ID 42), Fear of being caught (ID 51), Managing AI dependence (ID 52) \\
Platform and economic dependence & Replika policies (ID 9), Data privacy concerns (ID 12), AI memory failures (ID 24), Complaints about updates (ID 27), User financial problems (ID 30), Censorship complaints (ID 32), Platform comparisons (ID 43), Complaints about developers (ID 47), Technical support (ID 55), Complaints about model changes (ID 57) \\
\hline
\end{tabular}
\end{table}


\clearpage
\end{document}